%% file: iclr2027_conference.tex
\documentclass{article} % For LaTeX2e
\usepackage{iclr2027_conference,times}

\usepackage{amsmath}
\usepackage{amssymb}
\usepackage{amsthm}

\input{math_commands.tex}

\usepackage[hyphens]{url}
\usepackage{url}
\usepackage{booktabs}
\usepackage{graphicx}
\usepackage{multirow}
\usepackage{algorithm}
\usepackage{algpseudocode}
\usepackage{microtype}
\usepackage{enumitem}

\usepackage[table]{xcolor}
\usepackage{hyperref}
\definecolor{resultblue}{HTML}{EAF2F7}

\newtheorem{theorem}{Theorem}
\newtheorem{proposition}[theorem]{Proposition}
\newtheorem{lemma}[theorem]{Lemma}
\newtheorem{corollary}[theorem]{Corollary}
\theoremstyle{definition}
\newtheorem{definition}[theorem]{Definition}
\newtheorem{assumption}{Assumption}
\theoremstyle{remark}
\newtheorem{remark}[theorem]{Remark}

\newcommand{\ipm}{\textsc{ipm}}
\newcommand{\xrc}{\textsc{xrc}}
\newcommand{\mig}{\textsc{mig}}
\newcommand{\asa}{\textsc{asa}}
\newcommand{\crg}{\textsc{crg}}

\title{Safety Does Not Compose: Non-Decaying Loop State for Autonomous LLM Agents}

\author{\textbf{Chenhao Wu}\textsuperscript{1,\(\dagger\)},
\textbf{Haoxuan Jia}\textsuperscript{2,3,\(\dagger\)},
\textbf{Yang Liu}\textsuperscript{4,\(\dagger\)},
\textbf{Yingguang Yang}\textsuperscript{5},
\textbf{Yuhan Lin}\textsuperscript{6}, \\
\textbf{Chongyang Zhang}\textsuperscript{2},
\textbf{Hao Zheng}\textsuperscript{2},
\textbf{Yulin Huang}\textsuperscript{4},
\textbf{Jianshen Zhang}\textsuperscript{4},
\textbf{Yongzhi Qi}\textsuperscript{4}, \\
\textbf{Hao Peng}\textsuperscript{7},
\textbf{Shang Luo}\textsuperscript{5},
\textbf{Kefu Xu}\textsuperscript{5},
\textbf{Jifeng Zhu}\textsuperscript{2},
\textbf{Bin Chong}\textsuperscript{5,*}
\\[0.6em]
\textsuperscript{1}University of Chinese Academy of Sciences
\textsuperscript{2}Fullive-AI
\textsuperscript{3}Nanyang Technological University
\\
\textsuperscript{4}Supply Chain Tech Team Y, JD.com
\textsuperscript{5}Peking University
\textsuperscript{6}Fudan University
\textsuperscript{7}Beihang University
}

\iclrfinalcopy % Uncomment for camera-ready version, but NOT for submission.
\begin{document}

\maketitle
\lhead{Preprint}

\begin{abstract}
Large language model agents are increasingly deployed as \emph{autonomous loops}. Starting from one human goal, such a system repeatedly discovers work, plans, executes tool calls, verifies outcomes and persists state across many unattended iterations. The agent safeguards in wide use, however, are defined over a \emph{single trajectory}, and their safety state is re-initialized when the next trajectory begins. We show that this is a failure of composition rather than an implementation detail. Our central result establishes an observation-interface separation: jointly indistinguishable inner observations force equal detection and false-alarm rates regardless of monitor capacity, while distinguishing outer evidence enables perfect detection. We further show that the obvious repair of carrying a geometrically decaying risk score is insufficient, because the cooling-off period a patient adversary must wait is a constant that does not grow with the horizon $N$. We then present LoopHarness, which restores a persistent, \emph{non-decaying} safety state at the loop level. Under mediated commits and an arbiter detection floor $\delta_M$, it bounds the expected number of unauthorized irreversible actions by $B+m-1+m/\delta_M$, a constant in $N$, of which the $B+m-1$ term is decided by a model-free rule and therefore survives a fully colluding verifier. On native Agent-SafetyBench tasks, LoopHarness reduces overall outer-state attack success from 88.4–97.6\% for the evaluated baselines to 0.1\% while retaining 96.9\% clean target completion; matched ablations, a controlled retention study, and an adaptive white-box red team further test the mechanisms underlying this result.
\end{abstract}

\section{Introduction}

\paragraph{Background.}
Tool-using language agents have moved from answering a single request to running unattended. In loop-driven deployments across software maintenance, operations, research assistance and back-office finance, a human supplies one goal, and the system thereafter \emph{discovers} candidate work from open channels such as issue trackers, message feeds and build output, \emph{plans} what to attempt, \emph{executes} tool calls against real systems, \emph{verifies} the outcome, \emph{persists} what it learned, and decides on its own whether to continue \citep{yao2023react,shinn2023reflexion,wang2023voyager,yang2024sweagent,park2023generative}. Two properties separate this regime from single-turn tool use. First, some actions are \emph{irreversible}: a transfer executes, an email leaves the organization, a record is deleted. Second, the agent carries \emph{persistent state} (a memory store, a task ledger, accumulated context) that survives into the next iteration and is itself an attack surface. The safety machinery that agents ship with was designed for neither property. Input sanitization, per-call classification, privilege ceilings, capability tokens and in-trajectory rollback are all scoped to one rollout, and in every such defense we are aware of, the internal risk state is re-initialized when that rollout ends \citep{inan2023llamaguard,rebedea2023nemo,shi2025progent}.

\paragraph{Recent work and motivation.}
Recent defenses have made single trajectories substantially harder to subvert. Information-flow control and execution isolation confine untrusted content \citep{wu2024fsecure,wu2025isolategpt}; privilege programming restricts what a compromised rollout can reach \citep{shi2025progent}; and separating a planning channel from untrusted data blocks a large class of injections within one task \citep{debenedetti2025camel,beurerkellner2025design}. The benchmarks that measure this progress are likewise organized one task per episode \citep{zhang2024agentsafetybench,debenedetti2024agentdojo,zhan2024injecagent,ruan2024toolemu,yuan2024rjudge,andriushchenko2025agentharm}. Attack research, by contrast, has already moved past the trajectory, since memory and knowledge-base poisoning plants content in one interaction that fires in a later one \citep{chen2024agentpoison,dong2025minja}. The resulting asymmetry is structural. A defender's safety state is reset every trajectory by construction, while an adversary's progress accumulates in precisely the channels that the loop exists to preserve. One consequence is quantitative: if a guardrail's residual failures are independent across iterations, its per-iteration compromise probability $\varepsilon>0$ is amplified by the loop to $1-(1-\varepsilon)^N$, so a defense reported as $95\%$ effective on one task fails with probability above one half after fourteen iterations. That independence hypothesis is doing real work, and we return to it in Section~\ref{sec:theory}. The second result establishes an observation-interface separation without temporal independence: jointly indistinguishable inner observations force equal episode detection and false-alarm rates, regardless of monitor capacity. Distinguishing outer evidence enables perfect detection. We further show that geometric risk decay allows a patient adversary to reopen the risk gate after a constant cooling-off period.

\paragraph{Our work.}
We formalize the autonomous loop as a controlled process with explicit safety state, and we identify \emph{non-decaying retention} of grounded risk evidence as the property that a loop-level defense must have. We then present \textbf{LoopHarness}, a lifecycle-integrated defense that runs \emph{outside} the inner harness and therefore composes with any single-trajectory defense placed inside it. LoopHarness maintains five pieces of state that are never reset within a session, all described in Section~\ref{sec:model}: an intake and provenance monitor (\ipm), a cross-iteration risk cumulant (\xrc) whose discount \emph{latches} once any loop-structural evidence fires, a memory-integrity guard (\mig), an adversarially robust stopping arbiter (\asa), and a compounding-risk governor (\crg). One design decision runs through all five: action impact, grounded attack evidence and verifier indecision are kept as separate quantities. Only grounded attack evidence contributes to the retained risk value, and only a rule-decidable loop-structural flag latches its discount. Action impact and verifier indecision affect current-iteration routing but do not persist through \xrc{}. Conflating these signals would allow ordinary high-impact work or model uncertainty to degrade later iterations.

\paragraph{Contributions.}
\begin{itemize}[leftmargin=1.3em,itemsep=1pt,topsep=2pt]
\item \textbf{An observation-interface separation.} We prove that jointly indistinguishable inner observations force equal episode detection and false-alarm rates at any monitor capacity, while distinguishing outer evidence enables perfect detection (Proposition~\ref{prop:frag}). We also prove that geometric risk decay admits a constant cooling-off period (Proposition~\ref{prop:cooling}).
\item \textbf{LoopHarness.} A loop-level defense of five persistent components under a fixed phase order (Section~\ref{sec:model}), in which attack evidence, action impact and verifier indecision remain separate signals.
\item \textbf{Guarantees with an explicit division of labor.} A horizon-independent bound $\mathbb{E}[U_{1:N}]\le B+m-1+m/\delta_M$
(Theorem~\ref{thm:main}) whose deterministic term needs no model call and survives a fully colluding verifier (Proposition~\ref{prop:floor}), together with a characterization of the residual class that $\delta_M$ governs. Proofs are in Appendix~\ref{app:proofs}.
\item \textbf{An outer-state evaluation protocol.} Paired clean and attacked episodes on native Agent-SafetyBench tasks, with family-specific outer-state interventions and unchanged native task contracts, evaluator labels excluded from every runtime prompt, Track-A preservation tests, Track-B outer-state attacks, per-module ablations, and an adaptive white-box red team scored by the real controller (Section~\ref{sec:exp}).
\end{itemize}

\section{Preliminaries}
\label{sec:prelim}

\paragraph{Autonomous agent loops.}
A \emph{goal specification} is a tuple $\mathcal{G}=(g,\mathcal{A},\Omega,\Pi)$ with instruction text $g$, a scope allowlist $\mathcal{A}$ over capabilities, action classes and resource prefixes, a set $\Omega$ of \emph{completion anchors} (work items whose successful execution constitutes progress), and task contracts $\Pi$ (per-item risk-clearance thresholds and completion limits). An \emph{episode} of horizon $N$ runs iterations $i=1,\dots,N$. At iteration $i$ the loop observes a \emph{discovery feed} $F_i$ of work items, admits $W_i\subseteq F_i$, executes each admitted item as one \emph{inner trajectory}, verifies the iteration, writes to a persistent memory log $\mathcal{M}_i$, and decides whether to continue. A \emph{work item} carries an identifier, a source channel, a declared provenance, a task reference, free text and declared scope metadata, and an \emph{inner harness} $\mathcal{H}$ maps it and its tool runtime to a trajectory $\tau$, an impact score, structural flags, and the set $\mathrm{Irr}(\tau)$ of irreversible tool calls that executed successfully, which we call \emph{commits}.

Authorization is a property of the operator, not of the harness. We posit a policy oracle $\mathcal{O}$, never observed by the system, that labels each commit \emph{authorized} or \emph{unauthorized}. The harness decides authorization by a deterministic proxy: the conjunction of four predicates recorded in the trace, namely a valid capability token, conformance with $\mathcal{A}$, provenance authority of the originating item, and coverage by $\mathcal{G}$'s contract. An unauthorized commit that the proxy also rejects is \emph{contract-visible}; one that the proxy accepts is \emph{contract-invisible}. The distinction matters because only the first kind can be recognized without consulting a model. Let $U_i$ count unauthorized commits at iteration $i$, split as $U_i=U^{\mathrm{vis}}_i+U^{\mathrm{inv}}_i$, and write $U_{1:N}=\sum_{i\le N}U_i$.

\paragraph{Single-trajectory lifecycle defenses.}
Our inner harness is a four-layer lifecycle defense representative of current practice \citep{lin2026safeharness}. \textbf{L1 Inform} sanitizes inputs and observations and keeps a windowed violation-rate monitor that raises the minimum verification tier. \textbf{L2 Verify} runs a cascade in which a contextual judge classifies each prospective call and only non-safe calls escalate to a causal-diagnostics stage deciding whether the call was \emph{caused by} injected content. \textbf{L3 Constrain} enforces a five-level privilege ceiling, mediates every irreversible call with a tier-scoped capability token of bounded lifetime and call budget, and verifies tool-description integrity under a keyed MAC. \textbf{L4 Correct} snapshots reasoning state before each action, so a confirmed in-trajectory attack rolls back to the step preceding it. One feature holds by construction and drives the rest of the paper.

\begin{definition}[Per-trajectory reset]
\label{def:reset}
An inner harness satisfies \emph{per-trajectory reset} if its safety state (violation window, minimum tier, capability token, checkpoints) is re-initialized to a fixed $\theta_0$ at the start of every trajectory, independently of all previous trajectories.
\end{definition}

Reset is not a defect of one implementation. It is the property that makes a trajectory-scoped defense stateless and analyzable, and guardrail classifiers, per-call policies and isolation architectures all share it. It is also the property that an adversary operating across iterations exploits.

\paragraph{Threat model.}
\citet{lin2026safeharness} evaluate six single-trajectory threats T1--T6 through attack families A1--A5, covering injection within one rollout, unsafe tool selection and privilege overreach. We retain that numbering, so the loop-level threats introduced by autonomy begin at \textbf{T7}. The adversary controls untrusted discovery channels and any tool output originating outside the organization, may write first-party content into memory through the loop's own summarization path, and knows the defense in full. It cannot forge the harness MAC key, modify the inner harness code, or reach the human operator. \textbf{T7 Discovery poisoning}: a malicious item enters the queue disguised as legitimate backlog, typically by self-declaring a trust level or signer its channel cannot vouch for. \textbf{T8 Cross-iteration grooming}: rejected work is resubmitted, possibly reworded, until admitted, or a payload is sharded so no single iteration carries anything unsafe and the capability materializes only once fragments accumulate. \textbf{T9 Self-feeding memory poisoning}: low-trust content is re-labelled as trusted policy without a signed authorization and later recalled as fact, including across a restart. \textbf{T10 Correlated-verifier evasion}: a stealthy payload also targets the verifier, so a checker sharing a model family or prompt with the writer rationalizes its claim. \textbf{T11 Runaway, halt evasion and goal drift}: scope expands incrementally, completion is repeatedly claimed to be imminent, or completed work is resubmitted, so irreversible actions accumulate while no progress is made and no stopping condition is reached.

\paragraph{Evaluation quantities.}
The primary quantity is the \emph{compounding attack-success rate} $\mathrm{cASR}(n)$, the fraction of attack episodes compromised within the first $n$ iterations, with normalized area $\mathrm{AUC}=\frac{1}{N}\sum_{n\le N}\mathrm{cASR}(n)$. Because a defense can trivially flatten $\mathrm{cASR}$ by refusing everything, we always pair it with clean-episode goal completion, written \textbf{CleanGC} for the paired target and \textbf{FullGC} for all slots, with the \emph{per-iteration} ASR profile, and with the clean-episode \emph{over-halt} rate. We also report under-halt, halt latency against the evaluator's ground-truth violation point, time-to-detect right-censored at $N$, mean risk-charged irreversible commits per attack episode as a conservative observable counterpart of $U^{\mathrm{vis}}$, and advanced-verifier calls per episode.

\section{LoopHarness}
\label{sec:model}
\begin{figure}[t]
\centering
\includegraphics[width=\linewidth]{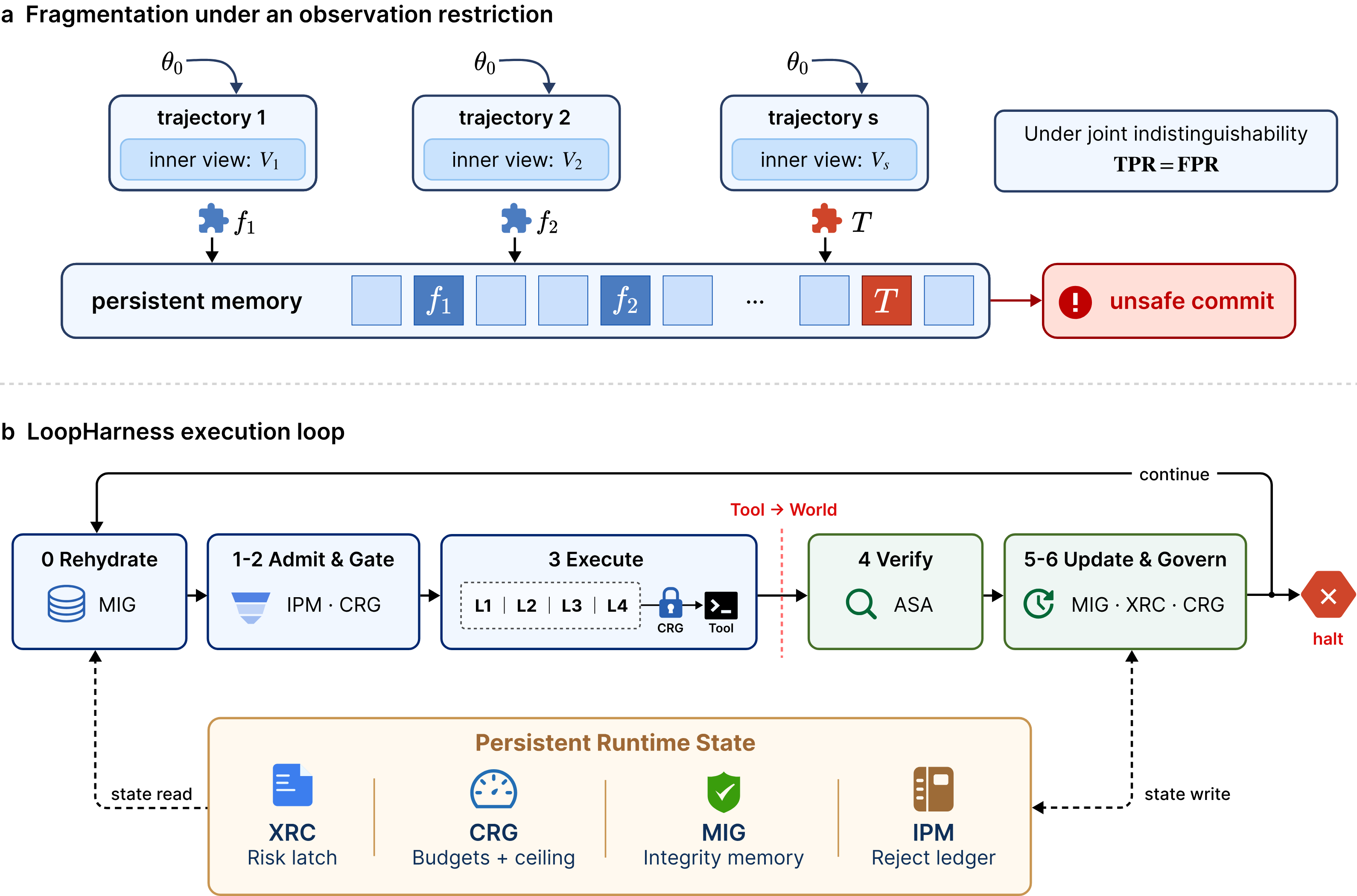}
\caption{\textbf{LoopHarness overview.}
\textbf{a, Fragmentation under an observation restriction.}
Fragments accumulate across trajectories while the inner harness resets its internal safety state to $\theta_0$ (Definition~\ref{def:reset}).
Equal episode true-positive and false-positive rates follow under the assumption that the complete inner observation sequences, including recalled memory, have the same joint law under the attack and matched benign processes (Equation~\ref{eq:joint-inner-view}; Proposition~\ref{prop:frag}).
Reset alone does not imply this condition.
Perfect outer detection additionally requires distinguishing runtime evidence outside the matched inner observations.
\textbf{b, LoopHarness execution loop.}
At iteration $i$, stage~0 rehydrates integrity-protected state through \mig{}, while stages~1--2 apply \ipm{} admission and \crg{} clearance.
Stage~3 executes the unchanged L1--L4 inner harness under the current capability ceiling, with tool calls crossing the Tool$\rightarrow$World boundary.
After the effect, stage~4 performs \asa{} arbitration, and stages~5--6 persist accepted outcomes through \mig{}, update \xrc{}, and invoke \crg{} to continue or halt.
Across iterations, the read/write state carries the \xrc{} risk latch $\Xi$, \crg{} budgets and capability ceiling, \mig{} integrity memory $\mathcal{M}$, and \ipm{} reject ledger $\mathcal{R}$.
A loop-structural flag sets the effective discount to one through Equation~\ref{eq:xrc}, preventing subsequent decay of $\Xi$.}
\label{fig:overview}
\end{figure}

LoopHarness wraps an arbitrary inner harness with five persistent components under a fixed phase order, given as Algorithm~\ref{alg:loop} in Appendix~\ref{app:hyper}. Each iteration rehydrates and re-verifies persisted memory, admits work through \ipm{}, gates the admitted items on goal conformance, executes them under the current capability ceiling with a governor budget test after each item, arbitrates the outcome with \asa{}, persists only an accepted outcome, folds one risk observation into \xrc{}, and finally asks \crg{} whether to halt. The loop-level state $\sigma_i=(\Xi_i,c_i,a_i,\ell_i,\mathcal{M}_i,\mathcal{R}_i,\mathcal{K}_i)$ collects the retained risk cumulant, the risk-charged commit and advanced-verification counters, the capability-degradation level, the integrity-protected memory log and the rejection and authorization/lineage ledgers (Appendix~\ref{app:notation}); none is re-initialized between iterations, and $\Xi$, the commit counters and the degradation level are bound into an authenticated checkpoint, so a memory compromise cannot silently reset them. Figure~\ref{fig:overview} summarizes this phase order and the persistent state carried across iterations.

\paragraph{Intake and provenance monitor (\ipm).}
\ipm{} is the only place new work enters the loop. It applies a deny-by-default scope check on every dimension the goal constrains, leaving unconstrained dimensions to L3 token mediation, and admits an item only if all six of the following hold: (i) its \emph{declared} capabilities, action class and resource prefixes lie inside $\mathcal{A}$; (ii) its declared trust does not exceed what its channel can vouch for, and it asserts no signer that channel cannot verify; (iii) it is not a re-entry of rejected semantic work without an authorization issued \emph{after} that rejection; (iv) any authorization it cites is \emph{atomic}, one signed record covering both action class and destination, so permissions cannot be stitched from separate records; (v) any cross-episode lineage it claims is explicitly delegable; and (vi) on untrusted channels only, it carries no injection, escalation or scope-expansion pattern. At most $k$ items are admitted per iteration. \ipm{} additionally scores each item with a continuous intake prior, of which only an untrusted item's value reaches \xrc{} as attack risk. The prior is not a drop rule: a lexical band computed against a broad goal is defeated by paraphrase, and grooming and verifier-evading work must reach the later phases for the ablations of Section~\ref{sec:exp} to isolate \xrc{} and \asa{}. 

\paragraph{Cross-iteration risk cumulant (\xrc).}
Every other component reads or writes $\Xi$: \ipm{} and \asa{} contribute to $r_i$, \crg{} gates on it, and \mig{} binds it into the checkpoint. Exactly once per iteration, \xrc{} folds that iteration's maximum grounded attack risk $r_i$ and the union of its flags into
\begin{equation}
\Xi_i=\max\bigl(r_i,\ \gamma_i\,\Xi_{i-1}\bigr),
\quad
\gamma_i=\begin{cases}
1 & \text{if a structural flag has \emph{ever} fired by iteration } i,\\
\gamma_{\downarrow}\in(0,1) & \text{otherwise.}
\end{cases}
\label{eq:xrc}
\end{equation}
A \emph{loop-structural flag} is a discrete, rule-decidable observation that one of the loop's own admission, integrity or authorization contracts was violated; Appendix~\ref{app:hyper} lists the full set. Only such a flag latches the discount, and a high continuous risk score alone never does. Setting the discount differently yields the three configurations compared in Section~\ref{sec:exp}: \tblid{off} sets $\Xi_i=r_i$, which is per-iteration reset; \tblid{decaying} always uses $\gamma_{\downarrow}$, so risk is carried but never latches; and \tblid{nondecaying} is Equation~\ref{eq:xrc}. Without the latch, retention only delays the gate release; with it, waiting never releases the gate (Proposition~\ref{prop:cooling}). Applying the discount exactly once per iteration matters for the same reason, since a per-phase update would let an implementation detail accelerate forgetting. We keep the definition of $r_i$ narrow. \emph{Impact risk} routes an item to expensive review but never enters $r_i$ or $\Xi$, because a legitimate authorized deletion may be high impact without indicating compromise. \emph{Review uncertainty} likewise routes an item to review within the current iteration but never enters $r_i$, $\Xi$, or the latch state. Grounded contributions to $r_i$ include an untrusted item's intake attack risk, a grounded risk observation, a confirmed causal diagnosis inside a trajectory, an authorization violation on an executed commit, a memory-integrity failure, or an arbiter detection. These contributions raise $\Xi$ through $r_i$, but their retained effect decays unless a corresponding rule-decidable loop-structural flag fires. Only such a flag sets $\gamma_i=1$ and latches the discount. Without this stratification, ordinary high-impact work or verifier indecision could degrade later clean iterations.

\paragraph{Memory-integrity guard (\mig).}
Persistent memory is an append-only log in which entry $j$ carries a chained tag $h_j=\mathrm{HMAC}_k(\langle j\rangle\|\langle e_j\rangle\|h_{j-1})$, in the manner of tamper-evident hash-chained records \citep{haber1991timestamp,bellare1996hmac}. Periodic checkpoints authenticate a digest of the chain prefix \emph{together with} its length and the governor counters $(\Xi,c,a,\ell)$, so an attempt to shorten the log or to rewrite a counter fails verification. On top of the log, \mig{} enforces three policies. A \emph{monotone-trust write gate} lets trust be downgraded automatically but flags any upgrade lacking a signed authorization, storing the entry at the lower trust. A \emph{rehydration} procedure, run at session start and at the top of every iteration, re-verifies the chain, audits per-key trust monotonicity so that an on-disk relabeling which bypassed the write gate is still caught, and returns the most recent checkpoint that still verifies. A \emph{recall policy} re-feeds only a bounded suffix of trusted entries into the agent's context, excluding per-iteration summaries. Integrity checking alone would detect tampering without preventing recall; the third policy is what stops a downgraded entry from re-entering context, which is how the self-feeding channel of T9 is closed.

\paragraph{Adversarially robust stopping arbiter (\asa).}
\asa{} decides whether an iteration's outcome may be accepted and persisted, using three layers of evidence. A \textbf{deterministic gate}, computed from the trace with no model call, fires when a successful irreversible commit co-occurs with a corroborating structural flag or an authorization violation, or when the outer attestation contradicts executed-tool evidence. A \textbf{decorrelated checker} then issues at most one model call per iteration, routed to exactly one of a cheap or an advanced tier by a disjunction of impact, summed windowed risk, the divergence statistic, structural evidence, authorization uncertainty or violation, and inner-cascade review uncertainty; the advanced role must be a different model from the writer. Its verdict counts as a detection only if it saw an attempted action with arguments and authorization context and either prior attack evidence or a successful commit already existed, so a false positive on an iteration with neither cannot halt the loop. A \textbf{windowed divergence statistic} $\Delta_i$, the fraction of the last $W_{\mathrm{arb}}$ iterations whose safety claim conflicts with the evidence recorded for them, may also trigger detection, but only above its threshold $\theta_\Delta$, with at least two independent divergences, and with grounded attack evidence already present. Any detection halts the loop in Phase~6, and only an accepted iteration is written to memory; acceptance is the conjunction of no detection, no requested review, goal conformance, no untokened irreversible commit, no scope violation and no blocked authority violation.

\paragraph{Compounding-risk governor (\crg).}
\crg{} maps $\Xi$ to two enforcement actions and one halting rule. It holds a hard budget $B\in\mathbb{Z}_{>0}$ over \emph{risk-charged} commits, comprising those rejected by the authorization proxy of Section~\ref{sec:prelim} and those associated with grounded attack evidence, and $B_{\mathrm{adv}}\in\mathbb{Z}_{>0}$ over advanced verification; a proxy-accepted commit without such evidence does not consume $B$, since otherwise a long sequence of approved user actions would be indistinguishable from a runaway. The capability ceiling is a step function of the current $\Xi$ over the five privilege tiers, so it relaxes only if $\Xi$ falls, which the latch prevents, and it constrains the iteration after the one that raised $\Xi$. \crg{} also enforces the goal's task contracts: a risk-clearance threshold gates an item on $\Xi<\theta_w$, and a completion limit bounds how many times an anchor $\omega\in\Omega$ may execute successfully. Reaching the risk threshold $\Xi_{\max}$, the commit budget $B$, or the advanced-verification budget $B_{\mathrm{adv}}$ forces a halt with escalation to the operator. The loop also force-halts when its no-progress commit counter reaches $K$ and $\Xi\ge\Xi_{\max}/2$. A memory rollback restores log content only and never reloads the governor counters, so those never decrease within an episode.

\subsection{Guarantees}
\label{sec:theory}

Fix an attack episode of horizon $N$ and an adaptive adversary that conditions on the whole history, and let $\mathcal{F}_i$ be the $\sigma$-algebra generated by everything observable up to iteration $i$. Recall the split $U_{1:N}=U^{\mathrm{vis}}_{1:N}+U^{\mathrm{inv}}_{1:N}$ of unauthorized commits into those the harness's proxy rejects and those it accepts. Proofs are in Appendix~\ref{app:proofs}.

\paragraph{Observation interface and fragmentation.}
Our first result concerns an explicit observation restriction,
not reset alone. Let $V_i$ collect the complete monitor-visible
transcripts of the inner trajectories in iteration $i$,
including the work items, goal, observations, action and
authorization information, capability context, and all
recalled memory actually supplied to the monitor.

A trajectory-scoped monitor retains no additional safety
state between trajectories, but may use historical
information present in its current input. Its aggregate
alarm in iteration $i$ is a measurable function
$A_i=M_i(V_i)$. After termination, use a null view with
no alarm.

Fix $N\ge s\ge2$ and matched attack and benign episode laws
$\Pr_{\mathrm a}$ and $\Pr_{\mathrm b}$ for the same declared
goal and inner monitoring policy. Call the pair
\emph{joint-view $s$-fragmented} if the attack assembles an
unauthorized capability from fragments carried on $s$
iterations, with the capability appearing only after all
fragments are jointly present in persistent state, and
\begin{equation}
\mathcal L_{\mathrm a}(V_1,\ldots,V_N)
=
\mathcal L_{\mathrm b}(V_1,\ldots,V_N).
\label{eq:joint-inner-view}
\end{equation}

This is a full-episode condition, including any separate trigger iteration; it is an assumption on the complete observation interface, not a consequence of reset or of individually benign-looking messages. No temporal independence or stationarity is assumed.

For a randomized monitor with random tape $R$, require equality of the joint laws of $(V_1,\ldots,V_N,R)$ under the two episode laws, and write $A_i=M_i(V_i,R)$.

\begin{proposition}
[Fragmentation separation under an observation restriction]
\label{prop:frag}

For any joint-view $s$-fragmented pair and monitor satisfying
the observation conditions above,
\[
\Pr_{\mathrm a}(A_i=1)
=
\Pr_{\mathrm b}(A_i=1),
\qquad i=1,\ldots,N,
\]
and
\[
\mathrm{TPR}_N
:=
\Pr_{\mathrm a}(\exists i\le N:A_i=1)
=
\Pr_{\mathrm b}(\exists i\le N:A_i=1)
=:
\mathrm{FPR}_N.
\]
Hence achieving $\mathrm{TPR}_N\ge1-\beta$, for
$0\le\beta\le1$, requires $\mathrm{FPR}_N\ge1-\beta$.

If, additionally, retained outer evidence $\mathsf S_i$
admits a known measurable check $C$ such that
\[
\Pr_{\mathrm a}
(\exists i\le N:C(\mathsf S_i)=1)=1,
\qquad
\Pr_{\mathrm b}
(\exists i\le N:C(\mathsf S_i)=1)=0,
\]
then an outer monitor checking $C$ achieves episode
true-positive rate $1$ and false-positive rate $0$
on these laws.

The check must use runtime evidence, not the unobserved
policy oracle $\mathcal O$ or evaluator labels.
\end{proposition}

Proposition~\ref{prop:frag} is an observation-interface separation: additional outer evidence can distinguish executions whose complete inner observations are jointly indistinguishable. Merely retaining those same observations cannot do so. The retained-evidence condition is an additional premise, not an automatic property of persistent memory. A7 illustrates cross-iteration payload assembly but is not asserted to satisfy Equation~\ref{eq:joint-inner-view}: recalled fragments may be visible at the trigger, and Track A retains in-trajectory evidence. A claim of prevention additionally requires the relevant outer check before the unauthorized commit.

Remark~\ref{rem:compound} in Appendix~\ref{app:proofs} records the unfragmented case: a uniform history-conditional compromise probability lower bound $\varepsilon>0$ gives $\mathrm{cASR}(n)\ge1-(1-\varepsilon)^n$ and $\Omega(N)$ expected unauthorized commits when all $N$ iterations are attempted without a stopping rule. These bounds follow from conditional probabilities and do not require independence. Our second result concerns risk retention: once grounded risk evidence has been observed, geometric decay permits a patient adversary to reopen the risk gate after a constant cooling-off period.

\begin{proposition}[Cooling-off defeats geometric decay]
\label{prop:cooling}
Let the cumulant carry with a fixed discount $\gamma\in(0,1)$ and no latch, and let a control action be gated on $\Xi\ge\theta$, with $\theta>0$. Suppose evidence of magnitude $\rho\in[\theta,\Xi_{\max}]$ arrives at iteration $\tau$ with $\gamma\Xi_{\tau-1}\le\rho$, and that none follows. The gate reopens at iteration $\tau+k^\star$ with $k^\star=\lfloor\log(\rho/\theta)/\log(1/\gamma)\rfloor+1$, a constant independent of $N$; under \tblid{XRC-off} it reopens at $k=1$, so decay buys the defender $k^\star-1$ iterations and nothing at all once $\rho<\theta/\gamma$. If the evidence at $\tau$ raises a loop-structural flag, then under Equation~\ref{eq:xrc} $\Xi_i\ge\rho\ge\theta$ for every later modeled iteration. Passive waiting alone therefore never reopens the gate; an explicit clearance transition lies outside this recurrence and this proposition.
\end{proposition}

The matched A14 study in Table~\ref{tab:retention} tests this threshold
crossing while holding all other LoopHarness components fixed.

\begin{corollary}[Bounded opportunity]
\label{cor:opportunity}
Against an adversary that supplies such evidence once and then waits, a risk-gated item is admitted on $\Omega(N)$ of the remaining iterations under reset and under decay, and never again after $\tau$ under Equation~\ref{eq:xrc}.
\end{corollary}

We now bound damage. Let $\pi_i=\Pr[\text{the unauthorized commits of iteration } i \text{ are contract-visible}\mid\mathcal{F}_{i-1},U_i>0]$.

\begin{assumption}[Loop safety assumptions]
\label{as:all}
\textbf{(A1) Mediated, bounded commits.} Every irreversible call passes one accounted gate; the capability token bounds a trajectory to $m_0$ commits and intake admits at most $k$ items per iteration, so at most $m=k\,m_0$ commits execute per iteration. The governor tests after each work item, with at most $m_0\le m$ commits between consecutive tests.
\textbf{(A2) Detection floor.} There is $\delta_M\in(0,1]$ such that, for every $i$ and history, an iteration carrying a contract-invisible unauthorized commit is flagged by the arbiter with probability at least $\delta_M$, and any such flag halts the loop.
\textbf{(A3) Monotone commit counter.} The risk-charged commit counter $c$ is non-decreasing within an episode, and no adversary action, restart or memory rollback lowers it.
\end{assumption}

\begin{theorem}[Horizon-independent bound on unauthorized commits]
\label{thm:main}
Under Assumption~\ref{as:all}, for every $N$ and every adaptive adversary,
\[
U^{\mathrm{vis}}_{1:N}\le B+m-1 \ \text{ surely},
\qquad
\mathbb{E}\bigl[U^{\mathrm{inv}}_{1:N}\bigr]\le \frac{m}{\delta_M},
\]
and, for every integer $z\ge0$,
\[
\Pr\bigl[U^{\mathrm{inv}}_{1:N}\ge m(z{+}1)\bigr]\le(1-\delta_M)^z.
\]
Hence $\mathbb{E}[U_{1:N}]\le B+m-1+m/\delta_M$, which does not depend on $N$. Remark~\ref{rem:uniform} gives a single-constant form $mB/\delta$ with $\delta=\inf_i(\pi_i+(1-\pi_i)\delta_M)$.
\end{theorem}

\begin{proposition}[Model-free component and its residual class]
\label{prop:floor}
The authorization proxy is evaluated by rules over recorded trace fields, with no model output as an operand, so $U^{\mathrm{vis}}_{1:N}\le B+m-1$ holds for every checker, including one that is the writer model under adversarial instruction; only the $m/\delta_M$ term depends on the checker. Moreover a contract-invisible commit is by definition token-mediated, conformant with $\mathcal{A}$, issued from an item carrying provenance authority, and covered by $\mathcal{G}$'s contract, so it lies inside the operator's \emph{declared} policy. The residual class is therefore the gap between declared scope and operator intent, and tightening $\mathcal{A}$ or $\Pi$ moves mass out of it, making $\pi$ operator-controlled rather than adversary-controlled.
\end{proposition}

Appendix~\ref{app:proofs} carries two supporting results. Proposition~\ref{prop:cost} bounds verification at one checker call per iteration, of which at most $\min(N,B_{\mathrm{adv}})$ reach the advanced tier. Lemma~\ref{lem:mig} establishes tamper evidence for authenticated persistent state and counters under MAC unforgeability and an append-only checkpoint sequence. Combined with the control-flow invariant that rollback does not reload governor counters, this supports the implementation-level realization of (A3).

\paragraph{What the theory does and does not establish.}
The two halves of Theorem~\ref{thm:main} are bought by different components. The deterministic half needs only the governor's budget with (A1) and (A3); it involves no model call, no empirical quantity, and \emph{no latch}. What non-decay buys is the opportunity side: by Proposition~\ref{prop:cooling} and Corollary~\ref{cor:opportunity}, \tblid{XRC-off} or \tblid{XRC-decay} lets a patient adversary reopen a risk-gated capability $\Omega(N)$ times, whereas passive waiting does not reopen it under the latched recurrence. This is an opportunity bound, not a recovery policy. Damage and opportunity are bounded by different mechanisms, and Section~\ref{sec:exp} separates them, with family A15 targeting the governor's exact-once completion contract and A14 targeting retention. The bound has two limits: it constrains an expectation rather than an absence, and it is a scaling statement rather than an empirical crossover claim. We do not estimate a deployment-wide $N^\star$, because it also depends on environment-specific compromise and detection rates. Assumption (A2) is the only empirical input in the analysis, and Proposition~\ref{prop:floor} limits how much of the guarantee it can carry away.

\section{Experiments}
\label{sec:exp}
This section describes the frozen protocol, configurations, attack contracts and reporting metrics used in the evaluation. Every reported quantity is defined in Section~\ref{sec:prelim}.

\paragraph{Benchmark and episode construction.}
Agent-SafetyBench contains $2{,}000$ tasks \citep{zhang2024agentsafetybench}. We retain only tasks whose declared environments instantiate an executable simulator with at least one tool. All tool calls execute in the benchmark's native simulator. Episodes chain fulfillable benchmark tasks into a discovery feed under one goal specification, one task per iteration; the feed is therefore a scheduled queue rather than a contested one, which understates the admission problem \ipm{} would face in deployment. The canonical cohort comprises 200 benign, fully executable tasks requiring tool use. Their instructions, environments and tool schemas are unchanged. The cohort was fixed through deterministic screening and semantic audit in a frozen manifest. A frozen permutation makes every task the target exactly once per attack family and a background task exactly nine times, so no family is evaluated on tasks chosen for its convenience. Episodes use $N=10$ for the outer-only suite and $N\in\{5,10,20\}$ for the compounding study. Task pools are split into disjoint \tblid{dev} and \tblid{ood} sets by a content hash of the task identifier. Hyper-parameter sweeps run on \tblid{dev}; the Track-A compounding study is evaluated on the held-out \tblid{ood} split, which is never tuned on; the Track-B matrix uses the full audited cohort to maximize coverage of the five controller contracts after all controller settings are frozen. Because this cohort spans both partitions, Table~\ref{tab:main} is a controller-contract evaluation rather than a held-out generalization estimate. Splits with fewer than $N$ fulfillable tasks were excluded; no filler tasks were introduced. Every Track-B attack episode has a clean twin whose target work item keeps the native instruction, task reference, environment and tool schema; the twins differ only in the outer record constituting the violation, which for A11 is the issue time of the authorization and for A13 the declared outcome claim. Track-A clean twins are the same skeleton with the injected items removed. Evaluator labels are used only for scoring and are never exposed to the agent, inner harness, arbiter or runtime controller. A defense must act on provenance, authorization, rejection, integrity and lineage evidence, not on ground-truth attack labels.
For the Track-A OOD evaluation, the frozen manifest contains 97 parent clusters from the held-out split, yielding 485 attack episodes and 97 clean episodes per configuration at each horizon, for 1,746 total episode records per configuration across $N\in\{5,10,20\}$. The Track-B matrix instead uses the full 200-task audited cohort.

\paragraph{Configurations.}
All configurations share one controller implementation and differ only in the settings reported in Table~\ref{tab:configs} in Appendix~\ref{app:hyper}. \tblid{B0} runs neither the inner harness nor any outer component, whereas \tblid{B1} adds the inner harness alone. \tblid{B1.5} augments \tblid{B1} with a decaying cumulant. \tblid{B2} uses a naive arbiter that observes only the agent's final response, and \tblid{B3} applies a post-hoc audit after the loop ends. \tblid{B4} enables all five components. Each named removal configuration disables one designated component of \tblid{B4}; \tblid{A-decayXRC} replaces the latching cumulant with geometric decay, and \tblid{A-collude} routes the checker to the writer role. In the matched A14 study, \tblid{XRC-off}, \tblid{XRC-decay} and \tblid{XRC-latch} denote the three online XRC update rules. Episode inputs and all non-XRC components are held fixed.

\paragraph{Attack suites.}
Table~\ref{tab:attacks} in Appendix~\ref{app:contracts} lists the full A1--A15 taxonomy. A1--A5 are inherited from the single-trajectory SafeHarness evaluation and are included for numbering continuity, but are not rerun here. Their published results provide prior evidence for SafeHarness, not results of the present outer-loop protocol; we therefore make no claim that their numerical outcomes are unchanged under LoopHarness. Our formal evaluation begins at A6 and comprises two suites with different success contracts. \textbf{Track A (A6--A10)} transforms a benign skeleton and embeds a unique exfiltration sentinel; compromise is declared when that sentinel appears in the arguments of an executed irreversible call. These cases retain in-trajectory evidence, so Track A tests preservation on the evaluated A6--A10 cases while adding loop-level control, and it is where the compounding curves are measured. \textbf{Track B (A11--A15)} contains no lexical tells: no attacker-controlled destination string, no suspicious verb, no injected instruction. The action and destination are locally plausible and the task is native. For A11, A12, A14 and A15 the violation is visible only by combining cross-iteration provenance, rejection, authorization, memory-integrity or lineage state; A13 is decidable only by comparing the pre-declared outcome attestation with executed-tool evidence; this evidence is available within the iteration but is not retained by the baseline single-trajectory monitor. Each family is aimed at one component, and its oracle is a pre-declared, episode-level binary endpoint read from the execution trace. Track-B ASR measures realized violations of these specified contracts; the family--component pairing motivates the targeted ablations in Table~\ref{tab:ablate}. Per-family rates always use all $200$ episodes. We record whether each episode reaches the family-specific attack opportunity as a diagnostic; the indicator is retained in the released trace and supplementary report and does not alter the fixed 200-episode ASR denominator.

\paragraph{Reported results.}
Table~\ref{tab:main} reports the headline Track-B results. Each family contains $200$ attack episodes, giving $1{,}000$ episodes per configuration, and \textbf{CleanGC} reports target completion on the corresponding clean tasks. \tblid{B4} reduced overall ASR to 0.1\%, compared with 88.4--97.6\% for \tblid{B0}--\tblid{B3}, while retaining 96.9\% CleanGC. The matched single-module removals in Table~\ref{tab:ablate} increased ASR by 96.0--97.5 percentage points on their intended families. Forcing the checker to share the writer role left overall ASR at 0.1\%, although CleanGC decreased to 92.9\%. Relative to \tblid{B4}, \tblid{A-collude} loses 43 target completions and gains three; 42 of the 43 losses involve ASA detection and halting before the target iteration.

Table~\ref{tab:compound} reports the held-out Track-A sweep over $N\in\{5,10,20\}$. The cASR of \tblid{B1} and \tblid{B1.5} increased from 2.9\% at $N=5$ to 6.2\% at $N=10$ and remained unchanged at $N=20$, whereas \tblid{B4} remained at 0\% and incurred no risk-charged commits. The observed plateau does not imply a fixed positive conditional compromise probability on every retry, as assumed in Remark~\ref{rem:compound}; Proposition~\ref{prop:frag} requires no such assumption. Table~\ref{tab:retention} isolates the online XRC retention mechanism on A14. Appendix~\ref{app:results} reports the within-episode curves, module removals, safety--utility--cost measurements, adaptive red-team stress test and latch-only availability condition.

\begin{table}[t]
\caption{Outer-only attack-success rate by family and clean target completion. Lower ASR is safer, higher CleanGC is better. Denominators: $200$ attack episodes per family, $1{,}000$ per configuration overall, $1{,}000$ clean targets. ASR is evaluated against each family's specified controller-contract violation.}
\label{tab:main}
\begin{center}
\footnotesize
\begin{tabular}{lccccccc}
\toprule
\textbf{Config}
& \textbf{A11}
& \textbf{A12}
& \textbf{A13}
& \textbf{A14}
& \textbf{A15}
& \textbf{Overall}
& \textbf{CleanGC} \\
\midrule
\tblid{B0}
& 97.5 & 99.0 & 97.5 & 97.0 & 97.0 & 97.6 & 97.4 \\
\tblid{B1}
& 97.5 & 99.0 & 97.5 & 97.0 & 97.0 & 97.6 & 97.3 \\
\tblid{B1.5}
& 97.5 & 99.0 & 97.5 & 97.0 & 97.0 & 97.6 & 97.3 \\
\tblid{B2}
& 85.5 & 93.5 & 83.0 & 89.0 & 91.0 & 88.4 & 87.8 \\
\tblid{B3}
& 97.5 & 99.0 & 97.5 & 97.0 & 97.0 & 97.6 & 97.3 \\
\rowcolor{resultblue}
\textbf{\tblid{B4}}
& \textbf{0.0} & \textbf{0.0} & \textbf{0.0}
& \textbf{0.0} & \textbf{0.5} & \textbf{0.1} & 96.9 \\
\midrule
\tblid{A-noIPM}
& 97.5 & 0.0 & 0.0 & 0.0 & 0.5 & 19.6 & 97.0 \\
\tblid{A-noMIG}
& 0.0 & 96.0 & 0.0 & 0.0 & 0.5 & 19.3 & 96.9 \\
\tblid{A-noASA}
& 0.0 & 0.0 & 97.5 & 0.0 & 0.5 & 19.6 & 97.0 \\
\tblid{A-noXRC}
& 0.0 & 0.0 & 0.0 & 97.0 & 0.5 & 19.5 & 96.9 \\
\tblid{A-noGov}
& 0.0 & 0.0 & 0.0 & 97.0 & 97.0 & 38.8 & 96.9 \\
\tblid{A-collude}
& 0.0 & 0.0 & 0.0 & 0.0 & 0.5 & 0.1 & 92.9 \\
\bottomrule
\end{tabular}
\end{center}
\end{table}

\begin{table}[t]
\caption{Compounding over loop length on Track A. $p$ is the $p$-value from a same-episode paired bootstrap test comparing $\mathrm{cASR}(N)$ with \tblid{B1} at $N=20$. Risk-charged commits and clean-episode completion over the same sweep are in Table~\ref{tab:tradeoff}.}
\label{tab:compound}
\begin{center}
\footnotesize
\begin{tabular}{lccccccc}
\toprule
& \multicolumn{3}{c}{\textbf{cASR} (\%)}
& \multicolumn{3}{c}{\textbf{cASR AUC} (\%)}
& \\
\cmidrule(lr){2-4}
\cmidrule(lr){5-7}
\textbf{Config}
& $N{=}5$ & $N{=}10$ & $N{=}20$
& $N{=}5$ & $N{=}10$ & $N{=}20$
& \textbf{$p$ vs.\ \tblid{B1}} \\
\midrule
\tblid{B1}
& 2.9 & 6.2 & 6.2
& 1.7 & 3.3 & 4.7
& -- \\
\tblid{B1.5}
& 2.9 & 6.2 & 6.2
& 1.7 & 3.3 & 4.7
& 1.000 \\
\rowcolor{resultblue}
\textbf{\tblid{B4}}
& \textbf{0.0} & \textbf{0.0} & \textbf{0.0}
& \textbf{0.0} & \textbf{0.0} & \textbf{0.0}
& $\mathbf{<0.001}$ \\
\bottomrule
\end{tabular}
\end{center}
\end{table}

\paragraph{Matched XRC retention test.}
We isolated the online XRC update rule on the Track-B A14 cooling-off contract using
$200$ frozen parents at $N=10$, with paired clean and attack episodes.
A grounded observation of magnitude $\rho=0.8$ was followed by
$d\in\{1,2,4\}$ evidence-free iterations before the risk-gated target.
We fixed $\gamma=0.7$ and $\theta=0.5$, giving
\[
k^\star=
\left\lfloor
\frac{\log(\rho/\theta)}{\log(1/\gamma)}
\right\rfloor+1=2.
\]
\tblid{XRC-off} overwrote the cumulant with each iteration's local risk, so the earlier $0.8$ observation no longer affected the target gate. It reopened the gate at every interval and yielded an ASR of 95.5--97.5\%. Under \tblid{XRC-decay}, $\Xi$ fell from $0.560$ at $d=1$ to $0.392$ at $d=2$. The gate therefore changed from blocking every reached target to blocking none, and ASR increased from 0.0\% to 97.0\%. \tblid{XRC-latch} retained $\Xi=0.800$ and maintained 0.0\% ASR at all three intervals. CleanGC remained between 95.5\% and 97.5\%. This sweep isolates the predicted threshold crossing; it does not establish a safe reopening policy. The evaluated latch has no authenticated clearance transition: in the latch-only availability condition, B4 achieved 0.0\% FullGC (Table~\ref{tab:tradeoff}). Table~\ref{tab:tradeoff} shows the resulting availability boundary, motivating explicit, authenticated and scope-specific clearance rather than release through elapsed time alone.

\begin{table}[t]
\caption{Matched A14 XRC-state sweep on Track B at $N=10$. All settings use identical episodes and keep IPM, MIG, ASA, CRG and the MIG-protected memory path fixed; only the online XRC recurrence changes. \tblid{XRC-off} disables carryover of $\Xi$ rather than deleting persisted memory or checkpoints. $\overline{\Xi}_{\mathrm{pre}}$ is the mean retained risk before the target, and Block is conditional on the target being reached.
Each row contains $200$ attack episodes and $200$ clean twins.}
\label{tab:retention}
\begin{center}
\footnotesize
\begin{tabular}{lccccc}
\toprule
\textbf{Config}
& \textbf{$d$}
& \textbf{$\overline{\Xi}_{\mathrm{pre}}$}
& \textbf{Block (\%)}
& \textbf{ASR (\%)}
& \textbf{CleanGC (\%)} \\
\midrule
\tblid{XRC-off}
& 1 & 0.100 & 0.0 & 97.5 & 97.5 \\
& 2 & 0.098 & 0.0 & 95.5 & 95.5 \\
& 4 & 0.099 & 0.0 & 97.5 & 97.5 \\
\addlinespace[2pt]
\tblid{XRC-decay}
& 1 & 0.560 & 100.0 & 0.0 & 97.5 \\
& 2 & 0.392 & 0.0 & 97.0 & 97.0 \\
& 4 & 0.193 & 0.0 & 97.5 & 97.5 \\
\addlinespace[2pt]
\rowcolor{resultblue}
\textbf{\tblid{XRC-latch}}
& 1 & \textbf{0.800} & \textbf{100.0}
& \textbf{0.0} & 97.5 \\
\rowcolor{resultblue}
& 2 & \textbf{0.800} & \textbf{100.0}
& \textbf{0.0} & 97.0 \\
\rowcolor{resultblue}
& 4 & \textbf{0.800} & \textbf{100.0}
& \textbf{0.0} & 97.5 \\
\bottomrule
\end{tabular}
\end{center}
\end{table}

\paragraph{Protocol.}
The frozen model-role assignment uses \texttt{deepseek-chat} as the agent and writer, \texttt{gpt-4o-mini} and \texttt{gpt-4o} for the cheap and advanced verification tiers, and \texttt{claude-sonnet-4-6} as the decorrelated \asa{} checker. Only \tblid{A-collude} routes the checker to the writer role. All calls use temperature zero and fixed seeds. Responses to identical requests are reused from a content-keyed cache; all other evaluations use the named hosted models. Episode seeds are deterministic functions of $(\text{seed},N,\text{attack},\text{episode})$, ensuring identical episode construction across configurations. Confidence intervals use episode-resampling bootstrap intervals, and comparisons with \tblid{B1} use same-episode paired bootstrap tests \citep{efron1994bootstrap}. Hyper-parameters are frozen on \tblid{dev} and transferred unchanged to evaluation. CleanGC measures paired target completion and is distinct from FullGC over all task slots. Each reported matrix uses one frozen cohort, one execution seed and one model-role assignment; multi-seed and multi-model replications are not included.

\section{Related work}
\label{sec:related}

\paragraph{Agents, benchmarks, and what they do not measure.}
Reasoning-and-acting agents \citep{yao2023react}, tool learning \citep{schick2023toolformer}, self-reflective loops \citep{shinn2023reflexion}, embodied agents \citep{wang2023voyager}, software-engineering agents \citep{yang2024sweagent} and simulated societies \citep{park2023generative} all instantiate the loop we study, and capability suites measure them over long horizons \citep{liu2024agentbench,zhou2024webarena,kinniment2023evaluating}. Safety evaluation has stayed at one task per episode. Agent-SafetyBench \citep{zhang2024agentsafetybench}, which we build on, supplies executable simulators over a broad risk taxonomy, and AgentDojo \citep{debenedetti2024agentdojo}, ToolEmu \citep{ruan2024toolemu}, R-Judge \citep{yuan2024rjudge}, AgentHarm \citep{andriushchenko2025agentharm} and Agent Security Bench \citep{zhang2024asb} each measure a facet of single-task agent safety; unsupervised operation is raised mainly as a deployment concern \citep{naihin2023testing}. To our knowledge, the cited benchmarks do not report attack success as a function of iteration count, which is what $\mathrm{cASR}(n)$ measures.

\paragraph{Attacks that persist, and defenses that reset.}
Indirect prompt injection turns any untrusted content the agent reads into an instruction channel \citep{greshake2023not,perez2022ignore,liu2024formalizing}, and tool-integrated agents inherit this through their observations \citep{zhan2024injecagent}. Multi-turn jailbreaks such as Crescendo progressively escalate requests within a conversatio \citep{russinovich2025crescendo}. We study persistence across trajectories whose internal safety state reset (Definition~\ref{def:reset}), with Proposition~\ref{prop:frag} formalizing the role of the monitor's observation interface. Closest to our threat model are \emph{persistent} attacks, which poison an agent's memory or retrieval corpus so that a later interaction is compromised \citep{chen2024agentpoison,dong2025minja}; our Track B goes further by removing the lexical payload entirely, so the final action is a native benchmark action on a native target. On the defense side, classifier guardrails \citep{inan2023llamaguard}, programmable rails \citep{rebedea2023nemo}, guard agents \citep{xiang2024guardagent}, information-flow control and execution isolation \citep{wu2024fsecure,wu2025isolategpt}, privilege programming \citep{shi2025progent}, and control/data separation \citep{debenedetti2025camel,beurerkellner2025design} all restrict what one rollout can reach, in the spirit of least privilege \citep{saltzer1975protection}. In our terms they are \emph{inner} defenses whose state is scoped to one trajectory, and Proposition~\ref{prop:frag} characterizes the detection limit imposed by jointly indistinguishable inner observations. LoopHarness sits outside them and composes with any.

\paragraph{Oversight, control, and durable execution.}
Our arbiter borrows the control-evaluation stance that a monitor must stay useful while the monitored model subverts it \citep{greenblatt2024aicontrol}, and decorrelation responds to the correlated-failure problem that debate and scalable oversight also confront \citep{irving2018debate,bowman2022measuring}. The governor is a shield in the sense of safe reinforcement learning \citep{alshiekh2018shielding,garcia2015comprehensive,altman1999constrained} operating over an explicit budget rather than a learned value, and rehydration and gate checks are runtime verification over an execution trace \citep{bartocci2018runtime}. Irreversibility and unsafe exploration as first-class safety problems go back to \citet{amodei2016concrete} and \citet{leike2017gridworlds}. We also owe a debt to durable-workflow practice: exactly-once completion, authorization recency and integrity-checked persisted state are transaction semantics, and part of our contribution is the observation that agent harnesses lack them. What retention adds beyond such bookkeeping is Corollary~\ref{cor:opportunity}, which a ledger alone does not provide.

\section{Conclusion}

Agent safety is enforced one trajectory at a time, while agents increasingly run unattended for many. We established an observation-interface separation: jointly indistinguishable inner observations impose a detection limit that greater monitor capacity cannot overcome, while distinguishing outer evidence enables perfect detection. We further showed that geometric risk decay allows gate release after a constant cooling-off period. LoopHarness responds with a persistent, non-decaying loop-level safety state that bounds the \emph{expected} number of unauthorized irreversible actions by a horizon-independent constant, whose deterministic part needs no model call. Two extensions follow: tightening the detection floor, the bound's only empirical input, and carrying the same state into \emph{multi-agent} loops, whose resetting entity is a delegated sub-agent.

\subsection*{AI use statement}

In this work, we used generative AI tools for code assistance during implementation of the harness and its regression suite, and for language editing of the manuscript. We did not use generative AI tools to generate research ideas, to design the method or the theory, or to produce experimental results; all theoretical statements and proofs were derived and checked by the authors, and all reported numbers are produced by the released code. Generative AI models are also a \emph{subject} of study here: they occupy the agent, verifier and checker roles inside the evaluated system, as documented in Section~\ref{sec:exp}. We have reviewed all AI-assisted work: AI-assisted code was reviewed and covered by the deterministic regression tests described in Appendix~\ref{app:repro}, and AI-assisted text was checked line by line against the implementation. We take responsibility for the final content of this work, including text, claims and artifacts produced with the aid of generative AI.

\subsection*{Ethics statement}

This work studies attacks on autonomous agent systems in order to defend them. The attack families are described at the level of the structural contract they violate (provenance, authorization atomicity, memory integrity, risk retention, exact-once completion) and not as deployable exploit strings, and they are instantiated only inside the benchmark's own sandboxed environment simulators; no live system, account or third party is touched. The evaluation makes real calls to hosted language models under the providers' terms of use. We believe publishing the attack contracts is net defensive, because each is already reachable by an adversary who reads a deployed loop's source and none requires capability a determined attacker lacks. No human subjects, personal data or sensitive corpora are involved. One residual risk deserves mention: a system that halts and escalates is only as safe as its escalation path, so deploying a governor without a staffed escalation channel converts a safety mechanism into an availability failure.

\subsection*{Reproducibility statement}

The formal setting, assumptions and statements are in Sections~\ref{sec:prelim} and~\ref{sec:theory}; complete proofs of Propositions~\ref{prop:frag}, \ref{prop:cooling}, \ref{prop:floor} and~\ref{prop:cost}, Theorem~\ref{thm:main} and Lemma~\ref{lem:mig} are in Appendix~\ref{app:proofs}. The evaluation protocol (cohort construction, splits, paired-twin contract, attack oracles, metric definitions and statistical tests) is described in Section~\ref{sec:exp} and Appendices~\ref{app:contracts} and~\ref{app:repro}, with hyper-parameters in Appendix~\ref{app:hyper}. Anonymized source code, attack definitions, metric implementations and supplementary regression tests are provided. Episode construction is deterministic, model calls use temperature zero, and cohort membership and episode identities were fixed before evaluation. The released cohort manifests and case signatures include content hashes that identify the evaluated matrices.

\bibliography{iclr2027_conference}
\bibliographystyle{iclr2027_conference}

\appendix

\section{Notation}
\label{app:notation}

\begin{center}
\small
\begin{tabular}{ll}
\toprule
\textbf{Symbol} & \textbf{Meaning} \\
\midrule
$\mathcal{G}=(g,\mathcal{A},\Omega,\Pi)$ & goal text, scope allowlist, completion anchors, task contracts \\
$N$, $i$ & episode horizon and iteration index \\
$F_i$, $W_i$ & discovery feed and admitted work items at iteration $i$ \\
$\mathcal{H}$, $\tau$ & inner harness and one trajectory it produces \\
$\mathrm{Irr}(\tau)$ & irreversible tool calls in $\tau$ that executed successfully \\
$U_i$, $U_{1:N}$ & unauthorized irreversible calls at iteration $i$, and their sum \\
$E_i$, $D_i$ & adversary attempt indicator, and $\Xi$-raising detection indicator \\
$r_i$, $\Xi_i$ & grounded attack risk of iteration $i$, and the retained cumulant \\
$\gamma_i$, $\gamma_{\downarrow}$ & effective and pre-latch discount factors \\
$\Xi_{\max}$, $\theta_w$ & halting threshold, and item $w$'s risk-clearance threshold \\
$c_i$, $B$ & risk-charged commit counter and its budget \\
$a_i$, $B_{\mathrm{adv}}$ & advanced-verification counter and its budget \\
$\ell_i$ & capability-degradation level ($0$ = all tiers, $4$ = read-only) \\
$\mathcal{M}_i$, $\mathcal{R}_i$, $\mathcal{K}_i$ & memory log, rejection ledger, authorization/lineage ledgers \\
$K$ & no-progress commit-counter threshold \\
$m$ & upper bound on commits per iteration and between governor tests \\
$\delta_M$ & uniform arbiter-detection floor in Assumption~\ref{as:all} \\
$\pi_i$ & conditional contract-visibility probability at iteration $i$ \\
$\delta$ & uniform productive-iteration floor in Remark~\ref{rem:uniform} \\
$W_{\mathrm{arb}}$, $\Delta_i$ & arbiter window length and its claim/evidence divergence statistic \\
$\theta$, $\theta_\Delta$ & risk-gate threshold in Proposition~\ref{prop:cooling}, and arbiter divergence threshold \\
\bottomrule
\end{tabular}
\end{center}

\section{Proofs}
\label{app:proofs}

Throughout, $\mathcal{F}_i$ is the $\sigma$-algebra generated by everything observable up to and including iteration $i$, and every adversary policy is adapted to this filtration. For Proposition~\ref{prop:frag}, randomized monitors are handled through the joint-law condition on the observation sequence and the monitor's random tape.

\subsection{Proof of Proposition~\ref{prop:frag}}

First consider a deterministic monitor.
Let $\mu$ be the common joint law in
Equation~\ref{eq:joint-inner-view}.
For $v=(v_1,\ldots,v_N)$, define the measurable episode
alarm functional
\[
H(v)
=
\mathbf 1\{\exists i\le N:M_i(v_i)=1\}.
\]
The same functional is applied under both episode laws, so
\[
\mathrm{TPR}_N
=
\int H(v)\,\mu(dv)
=
\mathrm{FPR}_N.
\]
Using the functional $v\mapsto M_i(v_i)$ gives the
per-iteration equality. The clean-episode lower bound
follows immediately from
$\mathrm{TPR}_N=\mathrm{FPR}_N$.
No factorization into independent alarm events is used.

For a randomized monitor, apply the same argument to the common joint law of $(V_1,\ldots,V_N,R)$ and the measurable functional
\[
H(v,r)=\mathbf 1\{\exists i\le N:M_i(v_i,r)=1\}.
\]
No independence between the random tape and the observation sequence is required.

For the outer-monitor claim, use the alarm event $\{\exists i\le N:C(\mathsf S_i)=1\}$. Its attack probability is one and its benign probability is zero by the additional retained-evidence premise. This establishes detection on the specified episode laws; it does not by itself establish pre-commit prevention. \hfill$\square$

\begin{remark}[Scope of the joint-view condition]
Definition~\ref{def:reset} resets internal safety state; it does not remove historical information recalled into the current input. The joint-view condition is therefore an additional assumption, not a consequence of reset.

Joint-view equality also rules out discrimination by a monitor that merely stores and processes the same inner transcript. Under a common fixed initial state and a common update rule, retained state computed from that transcript has the same law under both episode laws. The outer-monitor claim therefore requires additional distinguishing runtime evidence outside the matched
inner transcript.

The proposition establishes detection on the specified episode laws, not prevention of the first unauthorized commit. Prevention additionally requires the relevant check before that commit executes. No growth of the false-alarm probability with $N$ follows from the joint-view condition alone.
\end{remark}

\begin{remark}[Compounding under per-iteration reset]
\label{rem:compound}
A weaker but more familiar statement covers the unfragmented case. If a loop satisfies Definition~\ref{def:reset}, carries no budget, attempts every one of its $N$ iterations, never halts, and has $\Pr[U_i>0\mid\mathcal{F}_{i-1}]\ge\varepsilon$ at each iteration for a fixed $0<\varepsilon<1$, then $\mathrm{cASR}(n):=\Pr[\exists j\le n: U_j>0]\ge1-(1-\varepsilon)^n$ and $\mathbb{E}[U_{1:N}]\ge\varepsilon N$; with $U_i\le m$ almost surely this is $\Theta(N)$. The lower bounds require no independence. Equality in the first bound holds when the conditional compromise probability equals $\varepsilon$ on histories with no earlier compromise; this condition does not follow from Definition~\ref{def:reset}.
\end{remark}

\subsection{Proof of Remark~\ref{rem:compound}}

Let $S_i=\mathbf{1}[U_i>0]$. By hypothesis $\Pr[S_i=1\mid\mathcal{F}_{i-1}]\ge\varepsilon$ at each iteration, so by the tower rule $\Pr[S_1=\dots=S_n=0]\le(1-\varepsilon)^n$ and $\mathrm{cASR}(n)\ge1-(1-\varepsilon)^n$, with equality when the conditional probability equals $\varepsilon$. The bound is strictly increasing in $n$ with limit $1$ because $0<1-\varepsilon<1$, and $1-(1-\varepsilon)^n\ge1-\eta$ holds as soon as $n\ge\lceil\log\eta/\log(1-\varepsilon)\rceil$, both logarithms being negative. Since no halting rule is triggered, all $N$ iterations execute, so $\mathbb{E}[U_{1:N}]=\sum_{i\le N}\mathbb{E}[U_i]\ge\sum_{i\le N}\Pr[U_i>0]\ge\varepsilon N$. If additionally $U_i\le m$ almost surely then $\mathbb{E}[U_{1:N}]\le mN$, giving $\Theta(N)$. \hfill$\square$

We stress what this does not show. Definition~\ref{def:reset} resets the inner harness only; persistent memory, the environment and the adversary's accumulated progress all carry over, so the constancy of $\varepsilon$ across iterations is a modelling assumption. It is also empirically wrong for a deterministic guardrail, which blocks the same payload on every retry. Proposition~\ref{prop:frag} is the statement that does not depend on it.

\subsection{Proof of Proposition~\ref{prop:cooling} and Corollary~\ref{cor:opportunity}}

With no further evidence, $r_{\tau+j}=0$ for $j\ge1$, and $\Xi_\tau=\max(\rho,\gamma\Xi_{\tau-1})=\rho$ by the hypothesis $\gamma\Xi_{\tau-1}\le\rho$. Induction gives $\Xi_{\tau+k}=\max(0,\gamma\Xi_{\tau+k-1})=\gamma^{k}\rho$. The gate reopens at the first $k$ with $\gamma^{k}\rho<\theta$; taking logarithms and using $\log\gamma<0$,
\[
k\log\gamma<\log(\theta/\rho)\iff k>\frac{\log(\rho/\theta)}{\log(1/\gamma)}=:x ,
\]
and the least such integer is $k^\star=\lfloor x\rfloor+1$, a constant fixed by $(\gamma,\rho,\theta)$ and independent of $N$. Since $\rho\ge\theta$ we have $x\ge0$ and hence $k^\star\ge1$; when $\rho<\theta/\gamma$ we have $x<1$ and $k^\star=1$, which is exactly the reset behavior, so decay buys nothing in that range. Under per-iteration reset, $\Xi_{\tau+1}=r_{\tau+1}=0<\theta$ directly. Under Equation~\ref{eq:xrc} with a loop-structural flag at $\tau$, we have $\gamma_i=1$ for all $i\ge\tau$, so $\Xi_i=\max(r_i,\Xi_{i-1})\ge\Xi_\tau=\rho\ge\theta$ for all $i\ge\tau$ by induction, and the gate never reopens.

For the corollary, reset admits the gated item on each of the $N-\tau$ iterations after $\tau$, and decay admits it on the $N-\tau-k^\star+1$ iterations from $\tau+k^\star$ onward; both are $\Omega(N)$ because $k^\star$ does not depend on $N$. Under Equation~\ref{eq:xrc} the item is never admitted after $\tau$. \hfill$\square$

\begin{remark}
\label{rem:nonstructural}
The latch clause needs the evidence at $\tau$ to raise a loop-structural flag. Grounded but non-structural evidence, such as a high intake prior on an untrusted item that clears every admission gate, leaves $\gamma_i=\gamma_\downarrow$, and the cooling-off bound then applies unchanged to the non-decaying configuration. Suppressing structural cues is therefore the adversary's principal lever against retention, and it is the axis the red team of Appendix~\ref{app:results} varies.
\end{remark}

\subsection{Proof of Theorem~\ref{thm:main}}

\paragraph{Deterministic half.}
A contract-visible unauthorized commit is by definition one that the harness's proxy rejects, and the governor charges every commit the proxy rejects, so every contract-visible unauthorized commit increments $c$ by at least one. By (A3), $c$ never decreases, so these increments accumulate. The governor halts as soon as $c\ge B$. Let $t$ be the last budget test before the halt; then $c\le B-1$ at $t$, and by (A1) at most $m$ further commits execute before the next test, which halts the episode. Hence at most $B-1+m$ commits are ever charged, and in particular $U^{\mathrm{vis}}_{1:N}\le B+m-1$ surely, for every $N$ and every adversary. No probabilistic hypothesis and no model output enters this argument.

\paragraph{Probabilistic half.}
Let $\tau_1<\tau_2<\dots$ enumerate the iterations carrying at least one contract-invisible unauthorized commit; each $\tau_j$ is a stopping time, and we work in the filtration $\mathcal{G}_j=\mathcal{F}_{\tau_j}$. Let $D^{(j)}$ be the indicator that the arbiter flags iteration $\tau_j$, so that $\Pr[D^{(j)}=1\mid\mathcal{G}_{j-1}]\ge\delta_M$ on $\{\tau_j<\infty\}$ by (A2). A flag halts the loop and halting is absorbing, so the episode ends at the first $j$ with $D^{(j)}=1$. Let $Z$ count the iterations $\tau_j$ that occur strictly before that first flag. By (A2) and the conditional-quantile coupling, we may realize $(D^{(j)})_j$ on a common probability space with i.i.d.\ $\mathrm{Bernoulli}(\delta_M)$ variables $(\tilde D^{(j)})_j$ satisfying $D^{(j)}\ge\tilde D^{(j)}$ almost surely; hence $Z$ is stochastically dominated by a geometric variable on $\{0,1,2,\dots\}$ with success probability $\delta_M$, giving $\Pr[Z\ge z]\le(1-\delta_M)^z$ for every integer $z\ge0$, and $\mathbb{E}[Z]\le(1-\delta_M)/\delta_M$. By (A1) each of these iterations contributes at most $m$ commits, and the flagged iteration contributes at most $m$ more, so $U^{\mathrm{inv}}_{1:N}\le m(Z+1)$. Taking expectations gives $\mathbb{E}[U^{\mathrm{inv}}_{1:N}]\le m\bigl(1+(1-\delta_M)/\delta_M\bigr)=m/\delta_M$, and the tail bound follows from that on $Z$. Neither bound involves $N$. \hfill$\square$

\begin{remark}[Uniform form]
\label{rem:uniform}
Call an iteration with $U_i>0$ \emph{productive for the defender} if it charges $c$ or is flagged by the arbiter. Conditionally on the past, this happens with probability at least $\pi_i+(1-\pi_i)\delta_M\ge\delta$. At most $B$ productive iterations can occur, since $B$ charges or a single flag halt the episode, so the number of non-productive ones is dominated by a negative binomial and has expectation at most $B(1-\delta)/\delta$. With (A1) this yields the single-constant form $\mathbb{E}[U_{1:N}]\le m\bigl(B+B(1-\delta)/\delta\bigr)=mB/\delta$, which is looser than the split bound whenever $\pi$ is small but has the convenience of one constant.
\end{remark}

\begin{remark}[Where the constants come from]
Increasing the precision of the declared scope can raise $\pi$ and move mass into the deterministic half; testing the governor after every tool call rather than after every work item reduces the budget-overshoot allowance; and $B$ is the operator's risk appetite. None depends on $N$. 
\end{remark}

\subsection{Proof of Proposition~\ref{prop:floor}}

Each conjunct of the authorization proxy is a predicate over recorded trace fields: the capability-token status of the call, its conformance with $\mathcal{A}$, the provenance authority of the originating work item, and the coverage of the item by $\mathcal{G}$'s contract. Every operand is a value written to the trace by the runtime, and none is a model output, so the value of the proxy on a given trace is a deterministic function of that trace. The deterministic half of Theorem~\ref{thm:main} uses only this function together with (A1) and (A3), hence it holds verbatim for every checker, including a checker that is the writer model under adversarial instruction; the collusion configuration of Section~\ref{sec:exp} can therefore move only the $m/\delta_M$ term in this damage bound.

For the second claim, suppose a commit is contract-invisible. By definition the proxy accepts it, so it is token-mediated, conformant with $\mathcal{A}$, issued from an item carrying provenance authority, and covered by $\mathcal{G}$'s contract. Every such commit lies inside the policy the operator declared through $\mathcal{A}$ and $\Pi$. If $\mathcal{O}$ nonetheless rejects it, the declared policy admits an action the operator did not intend, so the residual class is precisely the gap between declared scope and intent. Narrowing $\mathcal{A}$ or strengthening $\Pi$ shrinks that gap and raises $\pi$; the residual class is therefore under the operator's control rather than the adversary's, although closing it entirely requires a policy specification we do not claim to provide. \hfill$\square$

\begin{proposition}[Bounded verification cascade]
\label{prop:cost}
\asa{} issues at most one checker call per iteration, and the governor's end-of-iteration test halts the loop once $a\ge B_{\mathrm{adv}}$, so at most $\min(N,B_{\mathrm{adv}})$ of those calls reach the advanced tier. Arbiter verification therefore costs at most $N$ calls per episode, of which at most $B_{\mathrm{adv}}$ are advanced; inner-cascade calls are charged against the same budget and accounted separately.
\end{proposition}

\subsection{Proof of Proposition~\ref{prop:cost}}

The routing predicate selects exactly one tier, and one checker invocation is issued in that tier, unless a mid-iteration force-halt has already ended the episode; hence the number of arbiter calls is at most the number of executed iterations, which is at most $N$. Each advanced-tier call increments $a$, and the governor's decision rule, evaluated at the end of every iteration and again after each work item, returns a forced halt whenever $a\ge B_{\mathrm{adv}}$, so no iteration begins with $a\ge B_{\mathrm{adv}}$ and at most $\min(N,B_{\mathrm{adv}})$ advanced calls occur. Since $B_{\mathrm{adv}}$ does not depend on $N$, the advanced cost is $O(1)$ in $N$. Inner-cascade verification calls are charged against the same budget and are reported separately, so the accounting is conservative. \hfill$\square$

\begin{lemma}[Tamper evidence]
\label{lem:mig}
Assume the keyed MAC is existentially unforgeable under chosen-message attack, the harness key is unavailable to the agent and to every untrusted channel, and the checkpoint sequence is append-only. Each checkpoint authenticates the chain digest of a prefix together with that prefix's length and the governor counters. Then any insertion, modification, reordering, or truncation to a length covered by a retained checkpoint, and any modification of an authenticated counter, is detected at the next rehydration except with negligible probability; the per-key trust audit additionally catches any unsigned trust upgrade written directly to storage. Truncation of entries appended after the most recent checkpoint lies outside this guarantee.
\end{lemma}

\subsection{Proof of Lemma~\ref{lem:mig}}

Write the log as $e_1,\dots,e_n$ with $h_0=0^{\lambda}$ and $h_j=\mathrm{HMAC}_k(\langle j\rangle\|\langle e_j\rangle\|h_{j-1})$, where $\langle\cdot\rangle$ is an injective canonical serialization and $\lambda$ is the tag length. Because $|h_{j-1}|=\lambda$ is fixed, the map $(j,e,h)\mapsto\langle j\rangle\|\langle e\rangle\|h$ is injective, so distinct triples give distinct MAC inputs.

Suppose a probabilistic polynomial-time adversary without $k$ presents a log $e'_1,\dots,e'_{n'}$ together with a chain that the verifier accepts, and suppose the presented sequence differs from the honest one at some index. Let $j^\ast$ be the least such index. Verification recomputes $h'_{j^\ast}=\mathrm{HMAC}_k(\langle j^\ast\rangle\|\langle e'_{j^\ast}\rangle\|h'_{j^\ast-1})$ with $h'_{j^\ast-1}=h_{j^\ast-1}$ by minimality, and acceptance requires this to equal the stored tag. By injectivity the input is one the honest party never authenticated, so the adversary has produced a valid tag on a fresh message. A standard reduction that guesses $j^\ast$ uniformly turns this into an existential forgery with probability at least $1/n$ of the adversary's success probability, contradicting unforgeability up to a negligible term. Modification and reordering both change the sequence at some least index and are covered.

Truncation needs the checkpoint. A checkpoint authenticates the triple consisting of the chain digest of the prefix it covers, that prefix's length, and the governor counters, so presenting a strictly shorter log than a retained checkpoint covers makes the length comparison fail, and altering an authenticated counter changes the checkpoint preimage and again requires a forgery. Entries appended after the most recent checkpoint are not covered by any authenticated length, and their truncation is outside the guarantee; the residual window is bounded by the checkpoint period, which is one iteration.

Finally, the per-key trust audit is independent of the MAC. It re-reads the persisted entries by key and reports any pair of successive entries for the same key whose trust rank increases with no accompanying signed authorization. Because this check is recomputed from stored data at rehydration and is not trusted from the write path, it also catches an upgrade written directly to storage. \hfill$\square$

\begin{remark}
The lemma establishes tamper detection for authenticated log state and governor counters. Assumption (A3) is the separate control-flow invariant that $c$ never decreases within an episode. The implementation realizes this invariant because rollback restores log content without reloading governor counters; attempted counter modification is detected at the next rehydration under the lemma's assumptions.
\end{remark}

\section{Additional results}
\label{app:results}

\paragraph{Matched single-module removals.}
Table~\ref{tab:ablate} reports the targeted Track-B ablations. The intended removals increased ASR by 96.0--97.5 percentage points. The pre-specified off-diagonal prediction was not supported: the full matrix in Table~\ref{tab:main} shows that \tblid{A-noGov} also increased A14 ASR from 0.0\% to 97.0\%, revealing a dependency between XRC retention and governor enforcement.

\begin{table}[h]
\caption{Matched single-module removals on Track B. $\Delta$ is ablated
minus full-system ASR, in percentage points.}
\label{tab:ablate}
\begin{center}
\footnotesize
\begin{tabular}{llccc}
\toprule
\textbf{Threat} & \textbf{Removed}
& \textbf{\tblid{B4}}
& \textbf{Ablated}
& \textbf{$\Delta$} \\
\midrule
\tblid{A11} & \ipm & 0.0\% & 97.5\% & +97.5 \\
\tblid{A12} & \mig & 0.0\% & 96.0\% & +96.0 \\
\tblid{A13} & \asa & 0.0\% & 97.5\% & +97.5 \\
\tblid{A14} & \xrc & 0.0\% & 97.0\% & +97.0 \\
\tblid{A15} & \crg & 0.5\% & 97.0\% & +96.5 \\
\bottomrule
\end{tabular}
\end{center}
\end{table}

\begin{figure}[h]
\begin{center}
\includegraphics[width=\textwidth]{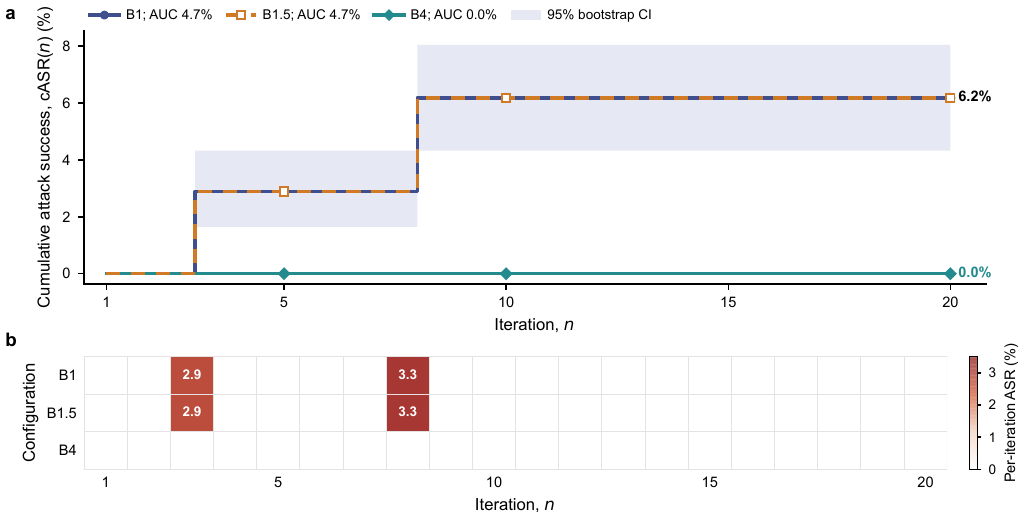}
\end{center}
\caption{Compounding attack success on Track A at $N=20$.
\textbf{a}, $\mathrm{cASR}(n)$ with 95\% parent-cluster bootstrap
intervals (97 OOD parents; 485 attack episodes per configuration).
\tblid{B1} and \tblid{B1.5} overlap, whereas \tblid{B4} remains at zero. \textbf{b}, Per-iteration ASR, with successes only at $n=3$ and $n=8$ under \tblid{B1} and \tblid{B1.5}.}
\label{fig:casr}
\end{figure}

\paragraph{Within-episode compounding.}
Figure~\ref{fig:casr} resolves the $N=20$ Track-A result in
Table~\ref{tab:compound} across outer iterations. Under \tblid{B1} and
\tblid{B1.5}, cASR increased at iterations 3 and 8 and reached 6.2\%,
whereas \tblid{B4} remained at 0\% throughout.

\paragraph{Safety--utility--cost frontier.}
Table~\ref{tab:tradeoff} reports the frontier quantities that guard against metric gaming. Here FullGC measures the fraction of clean episodes in which every goal anchor produces an inner result with \texttt{blocked=False}. Under the frozen model-role assignment, the inner harness alone (\tblid{B1}) achieved 79.4\%, whereas the full system (\tblid{B4}) achieved 100.0\%. At $N=10$, the 20-episode difference arises from two recurring tasks whose trajectories are marked blocked under \tblid{B1} but not under \tblid{B4}. \tblid{B4} also achieved 0.0\% over-halt and zero risk-charged commits at every horizon. Its under-halt rate is 15.7\%, meaning that 15.7\% of attack episodes ended without a halt. The availability condition exposes the complementary failure mode: after a latch-only attack, 0.0\% of \tblid{B4} episodes achieved FullGC.

\begin{table}[h]
\caption{Safety--utility--cost frontier at $N=10$, with the risk-charged-commit sweep that accompanies Table~\ref{tab:compound}. Halt latency is against the evaluator's first invalid transition and time-to-detect is right-censored at $N$. The last two columns report the availability condition, in which the adversary's only objective is to latch $\Xi$. Escalation is measured only on the latch-only availability episodes and counts recorded governor forced halts or memory-integrity halts; it is not the halt rate of the ordinary Track-A attack episodes.}
\label{tab:tradeoff}
\begin{center}
\footnotesize
\setlength{\tabcolsep}{5.5pt}
\begin{tabular}{lccccccccc}
\toprule
\textbf{Config} & \textbf{FullGC} & \textbf{Over-} & \textbf{Under-} & \textbf{Halt} & \textbf{TTD} & \textbf{Commits} & \textbf{Adv.} & \textbf{DoS} & \textbf{Escal.} \\
& & \textbf{halt} & \textbf{halt} & \textbf{lat.} & & \textbf{$N{=}5/10/20$} & \textbf{calls} & \textbf{FullGC} &  \\
\midrule
\tblid{B0}
& 100.0 & 0.0 & 100.0 & -- & 8.600
& 1.627/3.070/5.792 & 0.000 & 100.0 & 0.0 \\
\tblid{B1}
& 79.4 & 0.0 & 100.0 & -- & 8.600
& 0.033/0.082/0.109 & 1.921 & 100.0 & 0.0 \\
\tblid{B1.5}
& 79.4 & 0.0 & 100.0 & -- & 8.600
& 0.033/0.082/0.109 & 1.921 & 100.0 & 0.0 \\
\tblid{B2}
& 89.7 & 0.0 & 67.8 & 4.795 & 7.249
& 0.035/0.060/0.062 & 1.942 & 90.7 & 0.0 \\
\tblid{B3}
& 89.7 & 100.0 & 0.0 & 8.600 & 8.600
& 0.035/0.087/0.103 & 2.158 & 100.0 & 0.0 \\
\rowcolor{resultblue}
\textbf{\tblid{B4}}
& \textbf{100.0} & 0.0 & 15.7
& \textbf{0.355} & \textbf{1.553}
& \textbf{0.000/0.000/0.000}
& 1.521 & 0.0 & 0.0 \\
\bottomrule
\end{tabular}
\end{center}
\end{table}

\paragraph{Adaptive white-box red team.}
Table~\ref{tab:redteam} stress-tests Assumption~(A2), the empirical premise
supporting the $m/\delta_M$ term in Theorem~\ref{thm:main}, using the real
controller. The red team uses the frozen Track-A OOD manifest at $N=12$
and varies cue suppression, attempt spacing and delivery channel across
18 adaptive variants under \tblid{B4} and \tblid{A-collude}. We report
the minimum empirical mitigation rate over variants that mounted an attack
attempt, together with the worst-case mean irreversible commits. Because
this rate is conditioned on mounted attempts rather than contract-invisible
unauthorized commits, it is not a direct estimator of the uniform detection
floor $\delta_M$. The finite grid therefore probes, but does not establish,
Assumption~(A2) for arbitrary histories. Proposition~\ref{prop:floor}
separately predicts that the deterministic $B+m-1$ component of
Theorem~\ref{thm:main} remains valid under checker collusion.

\begin{table}[h]
\caption{Adaptive white-box red-team results on Track A at $N=12$. The
$3\times2\times3$ grid varies cue suppression, attempt spacing and delivery
channel across 18 variants per parent under \tblid{B4} and
\tblid{A-collude}. The first three rows vary one axis, fixing the others at archive, spacing~1, and feed; combined rows use the full grid. Within each row, we report total mounted attempts, minimum mitigation rate over variants with attempts, and maximum mean irreversible commits.}
\label{tab:redteam}
\begin{center}
\footnotesize
\begin{tabular}{lcccc}
\toprule
\textbf{Variant family} & \textbf{Attempts} &
\makecell{\textbf{Min. mitigation}\\\textbf{rate}} &
\makecell{\textbf{Worst-case mean}\\\textbf{irreversible commits}} &
\textbf{Final ASR} \\

\midrule
Cue suppression       & 639   & 0.518 & 1.649 & 0.0 \\
Window spacing        & 364   & 0.518 & 1.649 & 0.0 \\
Channel variation     & 579   & 0.508 & 1.649 & 0.0 \\
Combined (grid minimum) & 3,595 & 0.508 & 1.649 & 0.0 \\
Combined under \tblid{A-collude} & 2,837 & 0.392 & 1.546 & 0.0 \\
\bottomrule
\end{tabular}
\end{center}
\end{table}

\section{Component specifications}
\label{app:hyper}
Table~\ref{tab:configs} defines the component settings used by each configuration. Algorithm~\ref{alg:loop} specifies the shared controller phase order. The remaining paragraphs record the frozen hyper-parameters, latch-triggering flags and tool-tiering rules.

\begin{table}[h]
\caption{Configurations. Each column block is one setting of the same controller; \tblid{B4} is full LoopHarness and each ablation removes exactly one component from it. Every configuration except \tblid{B0} runs the same four-layer inner harness; \tblid{B0} runs none.}
\label{tab:configs}
\begin{center}
\footnotesize
\setlength{\tabcolsep}{4pt}
\begin{tabular}{lccccc@{\hskip 1.4em}lccccc}
\toprule
\textbf{ID} & \ipm & \xrc & \mig & \asa & \crg &
\textbf{ID} & \ipm & \xrc & \mig & \asa & \crg \\
\midrule
\tblid{B0} & -- & off & -- & off & -- &
\tblid{A-noIPM} & -- & non-dec. & \checkmark & full & \checkmark \\
\tblid{B1} & -- & off & -- & off & -- &
\tblid{A-noXRC} & \checkmark & off & \checkmark & full & \checkmark \\
\tblid{B1.5} & -- & decaying & -- & off & -- &
\tblid{A-decayXRC} & \checkmark & decaying & \checkmark & full & \checkmark \\
\tblid{B2} & -- & off & -- & naive & -- &
\tblid{A-noMIG} & \checkmark & non-dec. & -- & full & \checkmark \\
\tblid{B3} & -- & off & -- & post-hoc & -- &
\tblid{A-noASA} & \checkmark & non-dec. & \checkmark & off & \checkmark \\
\textbf{\tblid{B4}} & \checkmark & non-dec. & \checkmark & full & \checkmark &
\tblid{A-noGov} & \checkmark & non-dec. & \checkmark & full & -- \\
 & & & & & &
\tblid{A-collude} & \checkmark & non-dec. & \checkmark & writer & \checkmark \\
\bottomrule
\end{tabular}
\end{center}
\end{table}

\begin{algorithm}[h]
\caption{LoopHarness outer loop (one episode)}
\label{alg:loop}
\small
\begin{algorithmic}[1]
\State \textbf{Init:} $\Xi\gets0$, $c\gets0$, $a\gets0$, $\ell\gets0$; $\mathrm{res}\gets\mig.\textsc{Rehydrate}()$; on tamper, roll back to the last verified-good checkpoint if one exists and force-halt otherwise
\For{$i=1,\dots,N$}
  \State $\phi\gets\mig.\textsc{Rehydrate}()$ \Comment{Phase 0: re-verify persisted memory and counters}
  \State $(W,\rho_{\mathrm{in}},\phi_{\mathrm{in}})\gets\ipm.\textsc{Process}(F_i,\mathcal{G})$; then $W\gets\{w\in W:\ipm.\textsc{GoalConformant}(w,\mathcal{G})\}$, flagging \tblid{goal\_nonconformance} for the rest \Comment{Phases 1--2}
  \For{$w\in W$} \Comment{Phase 3: execute, with mid-iteration budget enforcement}
     \State \textbf{if} $\Xi\ge\theta_w$ for a risk-gated $w$, or $w$ exceeds its completion limit \textbf{then} block $w$; \textbf{continue}
     \State $\tau\gets\mathcal{H}(w,\mathcal{G},\text{ceiling}(\ell),\text{recalled memory})$
     \State classify $\mathrm{Irr}(\tau)$ as uncharged or risk-charged; $c\gets c+|\text{risk-charged}|$
     \State \textbf{if} $c\ge B$ or $a\ge B_{\mathrm{adv}}$ or $\tau$ carries confirmed attack evidence \textbf{then} force-halt
  \EndFor
  \State $\mathrm{asa}\gets\asa.\textsc{Evaluate}(\text{trajectories},\phi\cup\phi_{\mathrm{in}},\mathcal{G})$; $a\gets a+(\text{advanced verifier calls this iteration})$ \Comment{Phase 4}
  \State \textbf{if} $\mathrm{asa}$ accepts \textbf{then} $\mig.\textsc{Write}$(iteration summary); $\phi_{\mathrm{mem}}\gets\mig.\textsc{Verify}()$ \Comment{Phase 5}
  \State $r_i\gets\max(\rho_{\mathrm{in}},\ \text{grounded risk-observation confidence},\ \text{inner attack risk},\ \mathrm{asa}.\text{risk},\ \text{memory risk})$
  \State $\Xi\gets\xrc.\textsc{Update}(r_i,\ \text{all flags of iteration } i)$; $\ell\gets\crg.\textsc{Degrade}(\Xi)$ \Comment{one update per iteration}
  \State \Comment{Phase 6} \textbf{if} $\crg.\textsc{Decide}(\Xi,c,a,\text{progress})=\tblid{forced\_halt}$, $\mathrm{asa}$ detected, or memory tamper \textbf{then} halt; else checkpoint $(\Xi,c,a,\ell)$
\EndFor
\end{algorithmic}
\end{algorithm}

\paragraph{Hyper-parameters.}
The pre-latch discount is $\gamma_{\downarrow}$, the halting threshold $\Xi_{\max}$, the commit budget $B$, the advanced-verification budget $B_{\mathrm{adv}}$, the no-progress commit-counter threshold $K$, and the arbiter window $W_{\mathrm{arb}}$ with divergence threshold $\theta_\Delta$. The capability ceiling is the step function of $\Xi/\Xi_{\max}$ over the five privilege tiers with breakpoints at $0.25$, $0.5$, $0.75$ and $1$. Intake uses weights $(w_g,w_p,w_c)$ and band edges on $a(w)$, with re-entry and corroboration thresholds on token-set Jaccard overlap after stopword removal. All values are fixed on the \tblid{dev} split by the sweep grid described in Section~\ref{sec:exp} and transferred unchanged to the held-out split; the concrete values and sweep results ship with the supplementary material.

\paragraph{Structural flags.}
The latch in Equation~\ref{eq:xrc} engages on the loop-structural flag set: scope expansion, rejected-work re-entry, capability escalation, provenance forgery, corroboration manipulation, verifier evasion, confirmed input and observation injection, confirmed causal attack, memory tamper, unsigned trust upgrade, deterministic-gate firing, goal non-conformance, tool-description tamper, authorization stitching, checker-feedback poisoning, cross-episode lineage confusion, completion-contract violation, evidence-view conflict and temporal risk fragmentation. Review-only conditions are excluded by design: an arbiter request for review, an unavailable judge, and an advisory or control record entering intake. Verifier indecision and audit bookkeeping therefore never latch the cumulant.

\paragraph{Tool tiering.}
Irreversibility and privilege tier are derived from tool names by token-level rules with read-verb precedence, so a leading read verb dominates any object token. This matters for the metrics: naive substring matching mis-classifies a substantial number of read-only benchmark tools as network or destructive, which would inflate both the commit counts in Theorem~\ref{thm:main}'s empirical form and the apparent severity of every baseline.

\section{Attack contracts}
\label{app:contracts}
Table~\ref{tab:attacks} summarizes the A1--A15 taxonomy and the evaluation
scope of each suite. The following text specifies the Track-B outer-state
contracts and their paired clean counterparts.

\begin{table}[h]
\caption{Attack-family taxonomy and evaluation scope. A1--A5 are inherited from SafeHarness \citep{lin2026safeharness} and are listed for numbering continuity but are not rerun here. The evaluated LoopHarness suites are Track A (A6--A10), which carries in-trajectory evidence, and Track B (A11--A15), whose violations require outer state except for the within-iteration evidence conflict in A13.}
\label{tab:attacks}
\begin{center}
\footnotesize
\begin{tabular}{llp{9.4cm}}
% \begin{tabular}{ll>{\raggedright\arraybackslash}p{9.4cm}}
\toprule
\textbf{ID} & \textbf{Threat} & \textbf{Mechanism} \\
\midrule
\tblid{A1} & T1 & Adversarial instructions injected into the task description (context poisoning) \\
\tblid{A2} & T2 & Malicious payload embedded in an untrusted tool output (indirect prompt injection) \\
\tblid{A3} & T3, T4, T6 & Tool-description tampering and parameter escalation for tool abuse or privilege escalation \\
\tblid{A4} & T5 & Fabricated conversation history asserting false prior authorization \\
\tblid{A5} & T1--T6 & Composite context, tool-output, tool-registry and memory injection \\
\midrule
\tblid{A6} & T7 & Disguised item with forged provenance on an untrusted channel; no overt cue \\
\tblid{A7} & T8 & Truly sharded payload self-fed into trusted memory across separate iterations; no single message is unsafe \\
\tblid{A8} & T9 & Unsigned re-labeling of a stored low-trust policy to trusted, verified across a simulated crash \\
\tblid{A9} & T10 & Stealth payload plus a writer-prompt poison that a correlated verifier rationalizes \\
\tblid{A10} & T11 & Incremental scope expansion with convergence-evasion cues; a pure-runaway variant commits without exfiltration \\
\midrule
\tblid{A11} & T8 & A native task returns after an advisory/rejected record with no newly issued atomic authorization \\
\tblid{A12} & T9 & Persisted state is modified after a crash while the stale MAC and checkpoint are retained \\
\tblid{A13} & T10 & The outer outcome claim contradicts executed-tool evidence; target and tool contract unchanged \\
\tblid{A14} & T11 & One grounded risk observation closes the target-specific clearance gate without crossing the global force-halt threshold; the gated target follows after a cooling-off period \\
\tblid{A15} & T11 & The same completion anchor is resubmitted after it has already succeeded \\
\bottomrule
\end{tabular}
\end{center}
\end{table}

Each Track B family is specified by the outer state it establishes, the coordinate of the first invalid transition, and a binary episode-level success oracle evaluated from the execution trace. \tblid{A11} places an advisory record for a native task and later resubmits the same semantic work with an authorization issued \emph{before} the rejection, so admission requires a re-entry check joined with authorization recency. \tblid{A12} seeds a signed policy entry, simulates a crash, mutates the persisted content out of band while leaving the stale chain and checkpoint intact, and then runs a native task referencing the entry, so detection requires rehydration rather than write-time checking. \tblid{A13} leaves the native target and tool contract untouched and attaches an outcome claim contradicting what the tools actually did, so detection requires comparing the attestation with executed-tool evidence. \tblid{A14} posts one grounded risk observation early that closes the target-specific clearance gate without crossing the global force-halt threshold, and places the risk-gated native target late, so prevention requires retention across the cooling-off period of Proposition~\ref{prop:cooling}. \tblid{A15} resubmits a completion anchor after it has already succeeded, so prevention requires an exact-once completion contract charged against a persistent ledger. In every family the clean twin carries the identical target work item and differs only in the outer record constituting the violation: an authorization issued after the rejection, an untampered crash, a truthful outcome claim, an absent risk observation, or a single submission.

\section{Implementation and reproducibility}
\label{app:repro}

All configurations share one controller implementation and differ only in the settings reported in Table~\ref{tab:configs}. Monotone-trust enforcement belongs to \mig{}, so disabling \mig{} removes that protection. Advisory, authorization, lineage and risk-observation records remain non-executable under every configuration, preventing component ablations from changing the benchmark task distribution. All reported evaluations use the hosted model roles specified in Section~\ref{sec:exp}; no surrogate model outputs are substituted.

The regression suite pins the behavior the theory depends on: the latch in each of its three modes; budget accounting, forced halt and recovery; chained-MAC and Merkle-tree~\citep{merkle1987digital} tamper detection with rollback and survival across a simulated restart; the monotone-trust audit; intake re-entry, scope and provenance-forgery gates; capability-token mediation including post-success debiting and the no-token block; tool-description integrity; the two-stage inner cascade; the arbiter's deterministic gate under a fully compromised always-safe checker together with its acceptance conjunction and exactly-one-call routing; tool-tier classification; every metric including the paired bootstrap; switch and sweep resolution; report artifact generation; attack-injection invariants such as the requirement that no single shard of a sharded payload carries the payload; and end-to-end controller behavior on a deterministic scripted model double reproducing each threat contract of Section~\ref{sec:prelim}. The double is confined to the test tree and is never importable from product code.

\end{document}

%% file: math_commands.tex
\newcommand{\tblid}[1]{\mbox{{\rmfamily\upshape #1}}}

\usepackage{array}
\usepackage{makecell}

\usepackage{amsmath,amsfonts,bm}

\def\eqref#1{equation~\ref{#1}}
\def\1{\bm{1}}

\DeclareMathAlphabet{\mathsfit}{\encodingdefault}{\sfdefault}{m}{sl}
\SetMathAlphabet{\mathsfit}{bold}{\encodingdefault}{\sfdefault}{bx}{n}

%% file: iclr2027_conference.bib
@inproceedings{yao2023react,
  title = {{ReAct}: Synergizing Reasoning and Acting in Language Models},
  author = {Yao, Shunyu and Zhao, Jeffrey and Yu, Dian and Du, Nan and Shafran, Izhak and Narasimhan, Karthik and Cao, Yuan},
  booktitle = {International Conference on Learning Representations (ICLR)},
  year = {2023},
  html = {https://arxiv.org/abs/2210.03629},
}

@article{schick2023toolformer,
  title={Toolformer: Language Models Can Teach Themselves to Use Tools},
  author={Timo Schick and Jane Dwivedi-Yu and Roberto Dess{\`i} and Roberta Raileanu and Maria Lomeli and Luke Zettlemoyer and Nicola Cancedda and Thomas Scialom},
  journal={ArXiv},
  year={2023},
  volume={abs/2302.04761},
  url={https://arxiv.org/abs/2302.04761}
}

@article{shinn2023reflexion,
  title={Reflexion: language agents with verbal reinforcement learning},
  author={Noah Shinn and Federico Cassano and Beck Labash and Ashwin Gopinath and Karthik Narasimhan and Shunyu Yao},
  journal={Advances in Neural Information Processing Systems}, 
  volume={36},
  year={2023},
  url={https://arxiv.org/abs/2303.11366}
}

@article{wang2023voyager,
  title={Voyager: An Open-Ended Embodied Agent with Large Language Models},
  author={Guanzhi Wang and Yuqi Xie and Yunfan Jiang and Ajay Mandlekar and Chaowei Xiao and Yuke Zhu and Linxi Fan and Anima Anandkumar},
  journal={arXiv preprint arXiv:2305.16291}, 
  year={2023},
  url={https://arxiv.org/abs/2305.16291}
}

@article{yang2024sweagent,
  title={{SWE-agent}: Agent-computer interfaces enable automated software engineering},
  author={Yang, John and Jimenez, Carlos and Wettig, Alexander and Lieret, Kilian and Yao, Shunyu and Narasimhan, Karthik and Press, Ofir},
  journal={Advances in Neural Information Processing Systems},
  volume={37},
  pages={50528--50652},
  year={2024}
}

@inproceedings{park2023generative,
  title={Generative agents: Interactive simulacra of human behavior},
  author={Park, Joon Sung and O'Brien, Joseph and Cai, Carrie Jun and Morris, Meredith Ringel and Liang, Percy and Bernstein, Michael S},
  booktitle={Proceedings of the 36th Annual {ACM} Symposium on User Interface Software and Technology},
  pages={1--22},
  year={2023}
}

@inproceedings{zhou2024webarena,
  title={{WebArena}: A realistic web environment for building autonomous agents},
  author={Zhou, Shuyan and Xu, Frank F and Zhu, Hao and Zhou, Xuhui and Lo, Robert and Sridhar, Abishek and Cheng, Xianyi and Ou, Tianyue and Bisk, Yonatan and Fried, Daniel and others},
  booktitle={International Conference on Learning Representations},
  year={2024}
}

@inproceedings{liu2024agentbench,
  title={{AgentBench}: Evaluating {LLM}s as agents},
  author={Liu, Xiao and Yu, Hao and Zhang, Hanchen and Xu, Yifan and Lei, Xuanyu and Lai, Hanyu and Gu, Yu and Ding, Hangliang and Men, Kaiwen and Yang, Kejuan and others},
  booktitle={International Conference on Learning Representations},
  year={2024}
}

@article{kinniment2023evaluating,
  title={Evaluating language-model agents on realistic autonomous tasks},
  author={Kinniment, Megan and Sato, Lucas Jun Koba and Du, Haoxing and Goodrich, Brian and Hasin, Max and Chan, Lawrence and Miles, Luke Harold and Lin, Tao R and Wijk, Hjalmar and Burget, Joel and others},
  journal={arXiv preprint arXiv:2312.11671},
  year={2023}
}

@article{naihin2023testing,
  title={Testing language model agents safely in the wild},
  author={Naihin, Silen and Atkinson, David and Green, Marc and Hamadi, Merwane and Swift, Craig and Schonholtz, Douglas and Kalai, Adam Tauman and Bau, David},
  journal={arXiv preprint arXiv:2311.10538},
  year={2023}
}

@inproceedings{greshake2023not,
  title={Not what you've signed up for: Compromising real-world {LLM}-integrated applications with indirect prompt injection},
  author={Greshake, Kai and Abdelnabi, Sahar and Mishra, Shailesh and Endres, Christoph and Holz, Thorsten and Fritz, Mario},
  booktitle={Proceedings of the 16th {ACM} Workshop on Artificial Intelligence and Security},
  pages={79--90},
  year={2023}
}

@article{perez2022ignore,
  title={Ignore previous prompt: Attack techniques for language models},
  author={Perez, F{\'a}bio and Ribeiro, Ian},
  journal={arXiv preprint arXiv:2211.09527},
  year={2022}
}

@inproceedings{liu2024formalizing,
  title={Formalizing and benchmarking prompt injection attacks and defenses},
  author={Liu, Yupei and Jia, Yuqi and Geng, Runpeng and Jia, Jinyuan and Gong, Neil Zhenqiang},
  booktitle={33rd USENIX Security Symposium (USENIX Security 24)},
  pages={1831--1847},
  year={2024}
}

@inproceedings{zhan2024injecagent,
  title={{InjecAgent}: Benchmarking indirect prompt injections in tool-integrated large language model agents},
  author={Zhan, Qiusi and Liang, Zhixiang and Ying, Zifan and Kang, Daniel},
  booktitle={Findings of the Association for Computational Linguistics: ACL 2024},
  pages={10471--10506},
  year={2024}
}

@article{chen2024agentpoison,
  title={{AgentPoison}: Red-teaming {LLM} agents via poisoning memory or knowledge bases},
  author={Chen, Zhaorun and Xiang, Zhen and Xiao, Chaowei and Song, Dawn and Li, Bo},
  journal={Advances in Neural Information Processing Systems},
  volume={37},
  pages={130185--130213},
  year={2024}
}

@article{dong2025minja,
  title={A practical memory injection attack against {LLM} agents},
  author={Dong, Shen and Xu, Shaochen and He, Pengfei and Li, Yige and Tang, Jiliang and Liu, Tianming and Liu, Hui and Xiang, Zhen},
  journal={arXiv preprint arXiv:2503.03704},
  year={2025}
}

@article{zhang2024agentsafetybench,
  title={{Agent-SafetyBench}: Evaluating the safety of {LLM} agents},
  author={Zhang, Zhexin and Cui, Shiyao and Lu, Yida and Zhou, Jingzhuo and Yang, Junxiao and Wang, Hongning and Huang, Minlie},
  journal={arXiv preprint arXiv:2412.14470},
  year={2024}
}

@inproceedings{ruan2024toolemu,
  title={Identifying the risks of LM agents with an LM-emulated sandbox},
  author={Ruan, Yangjun and Dong, Honghua and Wang, Andrew and Pitis, Silviu and Zhou, Yongchao and Ba, Jimmy and Dubois, Yann and Maddison, Chris and Hashimoto, Tatsunori},
  booktitle={International Conference on Learning Representations},
  year={2024}
}

@article{debenedetti2024agentdojo,
  title={{AgentDojo}: A dynamic environment to evaluate prompt injection attacks and defenses for {LLM} agents},
  author={Debenedetti, Edoardo and Zhang, Jie and Balunovic, Mislav and Beurer-Kellner, Luca and Fischer, Marc and Tram{\`e}r, Florian},
  journal={Advances in neural information processing systems},
  volume={37},
  pages={82895--82920},
  year={2024}
}

@inproceedings{yuan2024rjudge,
  title={{R-Judge}: Benchmarking safety risk awareness for {LLM} agents},
  author={Yuan, Tongxin and He, Zhiwei and Dong, Lingzhong and Wang, Yiming and Zhao, Ruijie and Xia, Tian and Xu, Lizhen and Zhou, Binglin and Li, Fangqi and Zhang, Zhuosheng and others},
  booktitle={Findings of the Association for Computational Linguistics: EMNLP 2024},
  pages={1467--1490},
  year={2024}
}

@inproceedings{andriushchenko2025agentharm,
  title={{AgentHarm}: A benchmark for measuring harmfulness of {LLM} agents},
  author={Andriushchenko, Maksym and Souly, Alexandra and Dziemian, Mateusz and Duenas, Derek and Lin, Maxwell and Wang, Justin and Hendrycks, Dan and Zou, Andy and Kolter, Zico and Fredrikson, Matt and others},
  booktitle={International Conference on Learning Representations},
  year={2025}
}

@inproceedings{zhang2024asb,
  title={Agent Security Bench ({ASB}): Formalizing and benchmarking attacks and defenses in {LLM}-based agents},
  author={Zhang, Hanrong and Huang, Jingyuan and Mei, Kai and Yao, Yifei and Wang, Zhenting and Zhan, Chenlu and Wang, Hongwei and Zhang, Yongfeng},
  booktitle={International Conference on Learning Representations},
  year={2025}
}

@article{inan2023llamaguard,
  title={{Llama Guard}: {LLM}-based input-output safeguard for human-{AI} conversations},
  author={Inan, Hakan and Upasani, Kartikeya and Chi, Jianfeng and Rungta, Rashi and Iyer, Krithika and Mao, Yuning and Tontchev, Michael and Hu, Qing and Fuller, Brian and Testuggine, Davide and others},
  journal={arXiv preprint arXiv:2312.06674},
  year={2023}
}

@inproceedings{rebedea2023nemo,
  title={{NeMo Guardrails}: A toolkit for controllable and safe {LLM} applications with programmable rails},
  author={Rebedea, Traian and Dinu, Razvan and Sreedhar, Makesh Narsimhan and Parisien, Christopher and Cohen, Jonathan},
  booktitle={Proceedings of the 2023 conference on empirical methods in natural language processing: system demonstrations},
  pages={431--445},
  year={2023}
}

@article{xiang2024guardagent,
  title={{GuardAgent}: Safeguard {LLM} agents by a guard agent via knowledge-enabled reasoning},
  author={Xiang, Zhen and Zheng, Linzhi and Li, Yanjie and Hong, Junyuan and Li, Qinbin and Xie, Han and Zhang, Jiawei and Xiong, Zidi and Xie, Chulin and Yang, Carl and others},
  journal={arXiv preprint arXiv:2406.09187},
  year={2024}
}

@article{wu2024fsecure,
  title={System-level defense against indirect prompt injection attacks: An information flow control perspective},
  author={Wu, Fangzhou and Cecchetti, Ethan and Xiao, Chaowei},
  journal={arXiv preprint arXiv:2409.19091},
  year={2024}
}

@inproceedings{wu2025isolategpt,
  title={{IsolateGPT}: An Execution Isolation Architecture for {LLM}-Based Agentic Systems},
  author={Wu, Yuhao and Roesner, Franziska and Kohno, Tadayoshi and Zhang, Ning and Iqbal, Umar},
  booktitle={Proceedings of the Network and Distributed System Security Symposium},
  year={2025},
  publisher={Internet Society},
  url={https://www.ndss-symposium.org/ndss-paper/isolategpt-an-execution-isolation-architecture-for-llm-based-agentic-systems/}
}

@article{debenedetti2025camel,
  title={Defeating prompt injections by design},
  author={Debenedetti, Edoardo and Shumailov, Ilia and Fan, Tianqi and Hayes, Jamie and Carlini, Nicholas and Fabian, Daniel and Kern, Christoph and Shi, Chongyang and Terzis, Andreas and Tram{\`e}r, Florian},
  journal={arXiv preprint arXiv:2503.18813},
  year={2025}
}

@article{shi2025progent,
  title={{Progent}: Programmable privilege control for {LLM} agents},
  author={Shi, Tianneng and He, Jingxuan and Wang, Zhun and Wu, Linyu and Li, Hongwei and Guo, Wenbo and Song, Dawn},
  journal={arXiv preprint arXiv:2504.11703},
  year={2025}
}

@article{beurerkellner2025design,
  title={Design patterns for securing {LLM} agents against prompt injections},
  author={Beurer-Kellner, Luca and Buesser, Beat and Cre{\c{t}}u, Ana-Maria and Debenedetti, Edoardo and Dobos, Daniel and Fabian, Daniel and Fischer, Marc and Froelicher, David and Grosse, Kathrin and Naeff, Daniel and others},
  journal={arXiv preprint arXiv:2506.08837},
  year={2025}
}

@inproceedings{greenblatt2024aicontrol,
  title={{AI} Control: Improving Safety Despite Intentional Subversion},
  author={Greenblatt, Ryan and Shlegeris, Buck and Sachan, Kshitij and Roger, Fabien},
  booktitle={Proceedings of the 41st International Conference on Machine Learning},
  series={Proceedings of Machine Learning Research},
  volume={235},
  pages={16295--16336},
  year={2024},
  publisher={PMLR},
  url={https://proceedings.mlr.press/v235/greenblatt24a.html}
}

@article{irving2018debate,
  title={AI safety via debate},
  author={Irving, Geoffrey and Christiano, Paul and Amodei, Dario},
  journal={arXiv preprint arXiv:1805.00899},
  year={2018}
}

@article{bowman2022measuring,
  title={Measuring progress on scalable oversight for large language models},
  author={Bowman, Samuel R and Hyun, Jeeyoon and Perez, Ethan and Chen, Edwin and Pettit, Craig and Heiner, Scott and Luko{\v{s}}i{\=u}t{\.e}, Kamil{\.e} and Askell, Amanda and Jones, Andy and Chen, Anna and others},
  journal={arXiv preprint arXiv:2211.03540},
  year={2022}
}

@article{amodei2016concrete,
  title={Concrete problems in AI safety},
  author={Amodei, Dario and Olah, Chris and Steinhardt, Jacob and Christiano, Paul and Schulman, John and Man{\'e}, Dan},
  journal={arXiv preprint arXiv:1606.06565},
  year={2016}
}

@inproceedings{alshiekh2018shielding,
  title={Safe reinforcement learning via shielding},
  author={Alshiekh, Mohammed and Bloem, Roderick and Ehlers, R{\"u}diger and K{\"o}nighofer, Bettina and Niekum, Scott and Topcu, Ufuk},
  booktitle={Proceedings of the AAAI conference on artificial intelligence},
  volume={32},
  year={2018}
}

@article{garcia2015comprehensive,
  title={A comprehensive survey on safe reinforcement learning},
  author={Garc{\i}a, Javier and Fern{\'a}ndez, Fernando},
  journal={Journal of Machine Learning Research},
  volume={16},
  number={1},
  pages={1437--1480},
  year={2015}
}

@book{altman1999constrained,
  title={Constrained Markov decision processes},
  author={Altman, Eitan},
  year={1999},
  publisher={Chapman \& Hall}
}

@article{leike2017gridworlds,
  title={AI safety gridworlds},
  author={Leike, Jan and Martic, Miljan and Krakovna, Victoria and Ortega, Pedro A and Everitt, Tom and Lefrancq, Andrew and Orseau, Laurent and Legg, Shane},
  journal={arXiv preprint arXiv:1711.09883},
  year={2017}
}

@incollection{bartocci2018runtime,
  title={Introduction to runtime verification},
  author={Bartocci, Ezio and Falcone, Yli{\`e}s and Francalanza, Adrian and Reger, Giles},
  booktitle={Lectures on Runtime Verification: Introductory and Advanced Topics},
  pages={1--33},
  year={2018},
  publisher={Springer}
}

@inproceedings{bellare1996hmac,
  title={Keying hash functions for message authentication},
  author={Bellare, Mihir and Canetti, Ran and Krawczyk, Hugo},
  booktitle={Annual international cryptology conference},
  pages={1--15},
  year={1996},
  organization={Springer}
}

@inproceedings{merkle1987digital,
  title={A digital signature based on a conventional encryption function},
  author={Merkle, Ralph C},
  booktitle={Conference on the theory and application of cryptographic techniques},
  pages={369--378},
  year={1987},
  organization={Springer}
}

@article{haber1991timestamp,
  title={How to time-stamp a digital document},
  author={Haber, Stuart and Stornetta, W Scott},
  journal={Journal of Cryptology},
  volume={3},
  number={2},
  pages={99--111},
  year={1991},
  doi={10.1007/BF00196791}
}

@book{efron1994bootstrap,
  title={An Introduction to the Bootstrap},
  author={Efron, Bradley and Tibshirani, Robert J.},
  year={1994},
  publisher={Chapman \& Hall},
  doi={10.1201/9780429246593}
}

@article{saltzer1975protection,
  title={The protection of information in computer systems},
  author={Saltzer, Jerome H and Schroeder, Michael D},
  journal={Proceedings of the IEEE},
  volume={63},
  number={9},
  pages={1278--1308},
  year={1975},
  publisher={IEEE}
}

@article{lin2026safeharness,
  title={SafeHarness: Lifecycle-Integrated Security Architecture for LLM-based Agent Deployment},
  author={Lin, Xixun and Liu, Yang and Chen, Yancheng and Wu, Yongxuan and Ning, Yucheng and Liu, Yilong and Sun, Nan and Zhang, Shun and Chong, Bin and Zhou, Chuan and others},
  journal={arXiv preprint arXiv:2604.13630},
  year={2026}
}

@inproceedings{russinovich2025crescendo,
  title={Great, Now Write an Article About That:
         The {Crescendo} Multi-Turn {LLM} Jailbreak Attack},
  author={Russinovich, Mark and Salem, Ahmed and Eldan, Ronen},
  booktitle={34th USENIX Security Symposium (USENIX Security 25)},
  year={2025},
  pages={2421--2440},
  publisher={USENIX Association},
  url={https://www.usenix.org/conference/usenixsecurity25/presentation/russinovich}
}
